\RequirePackage{rotating}
\documentclass[fleqn,usenatbib]{mnras}
\usepackage{comment}
\usepackage[normalem]{ulem}
\usepackage{gensymb}
\usepackage{url}
\usepackage{caption}
\usepackage{newtxtext,newtxmath}
\usepackage{adjustbox}
\newcommand{\HI}{H\,{\sc i}}

\newcommand{\kms}{~km\,s$^{-1}$}
\usepackage[autostyle]{csquotes}
\usepackage{textcomp}
\usepackage{siunitx}
\usepackage{rotating}
\usepackage{float}
\MakeRobust{\overrightarrow}
\usepackage{hyperref}
\usepackage[bigfiles]{media9}

\usepackage[T1]{fontenc}
\usepackage{ae,aecompl}
\usepackage{rotating}

\usepackage{graphicx}	
\usepackage{subfig}
\usepackage{amsmath}	
\usepackage{adjustbox}
\usepackage{xcolor}
\usepackage{ulem}

\usepackage[normalem]{ulem}
\newcommand\ianjaout{\bgroup\markoverwith
{\textcolor{red}{\rule[.5ex]{3pt}{0.4pt}}}\ULon}

\title[]{MIGHTEE-\HI: Possible interactions with the galaxy NGC~895}
\author[Namumba et al.]
{B.\ Namumba$^{1,2,3}$\thanks{E-mail: bnamumba@gmail.com},
J.\ Román$^{2,4,5}$,
J. Falcón-Barroso$^{2,4}$,
J.\ H.\ Knapen$^{2,4}$,
R.\ Ianjamasimanana$^{6}$,
E.\ Naluminsa$^{7}$,
\newauthor
G.\ I.\ G.\ Józsa$^{8,3}$,
M.\ Korsaga$^{9,10}$,
N.\ Maddox$^{11,12}$,
B.\ Frank$^{13,14,15}$,
S.\ Sikhosana$^{16}$,
S.\ Legodi$^{13}$,
C.\ Carignan$^{15,17,10}$,
\newauthor
A.\ A.\ Ponomareva$^{18}$,
T.\ Jarrett$^{15}$,
D.\ Lucero $^{19}$,
O.\ M.\ Smirnov$^{3,13}$,
J.\ M.\ van der Hulst$^{5}$,
D.\ J.\ Pisano$^{15}$,
\newauthor
K.\ Malek$^{20,21}$,
L.\ Marchetti$^{15,14,22}$,
M.\ Vaccari$^{15,14,22}$,
M. \ Jarvis$^{18,23}$,
M. \ Baes$^{24}$,
M. \ Meyer$^{25,26}$,
\newauthor
E.\ A.\ K.\ Adams $^{26,5}$
H. \ Chen$^{27}$,
J.\ Delhaize$^{15}$,
S.\ H.\ A. \ Rajohnson$^{15}$,
S. \ Kurapati$^{15}$,
I.\ Heywood$^{18,3,13}$,
\newauthor
L.\ Verdes-Montenegro$^{6}$
\\
$^1$Wits Centre for Astrophysics, School of Physics, University of the Witwatersrand, 1 Jan Smuts Avenue, 2000, South Africa\\
$^2$Instituto de Astrof\'isica de Canarias, c/ V\'ia L\'actea s/n, 38205 La Laguna, Tenerife, Spain \\ 
$^3$Department of Physics and Electronics, Rhodes University, PO Box 94, Makhanda, 6140, South Africa \\ 
$^4$Departamento de Astrof\'isica, Universidad de La Laguna, E-38200 La Laguna, Spain \\
$^{5}$ Kapteyn Astronomical Institute, University of Groningen, PO Box 800, 9700 AV Groningen, The Netherlands\\ 
$^6$Instituto de Astrofísica de Andalucía (CSIC), Glorieta de la Astronomía, E-18008 Granada, Spain \\
$^7$South African Astronomical Observatory, P.O. Box 9, Observatory 7935, South Africa \\
$^8$Max-Plank-Institut f\"ur Radioastronomie, Auf dem H\"ugel 69, 53121 Bonn, Germany\\
$^9$Universit\'e de Strasbourg, CNRS, Observatoire astronomique de Strasbourg, UMR 7550, F-67000 Strasbourg, France\\
$^{10}$Observatoire d'Astrophysique de l'Université Ouaga I Pr Joseph Ki-Zerbo (ODAUO), BP 7021, Ouaga 03, Burkina Faso \\
$^{11}$School of Physics, H.H. Wills Physics Laboratory, Tyndall Avenue, University of Bristol, Bristol, BS8 1TL, UK\\
$^{12}$University Observatory, Faculty of Physics, Ludwig-Maximilians-Universit\"at, Scheinerstr. 1, 81679 Munich, Germany\\
$^{13}$South African Radio Astronomy Observatory, 2 Fir Street, Black River Park, Observatory, Cape Town, 7925, South Africa\\
$^{14}$The Inter-University Institute for Data Intensive Astronomy (IDIA), and University of Cape Town, Private Bag X3, Rondebosch, 7701, South Africa\\
$^{15}$Department of Astronomy, University of Cape Town, Private Bag X3, Rondebosch 7701, South Africa\\
$^{16}$School of Mathematics, Statistics $\&$ Computer Science, University of KwaZulu-Natal, Westville Campus,
Durban 4041, South Africa\\
$^{17}$Département de physique, Université de Montréal, Complexe des sciences MIL, 1375 Avenue Thérèse-Lavoie-Roux, 
Montréal, Qc, Canada H2V 0B3 \\
$^{18}$Oxford Astrophysics, Denys Wilkinson Building, University of Oxford, Keble Rd, Oxford, OX1 3RH, UK \\
$^{19}$Department of Physics, Virginia Polytechnic Institute and State University, 50 West Campus Drive, Blacksburg, VA 24061, USA \\
$^{20}$National Centre for Nuclear Research, ul. Pasteura 7, 02-093 Warsaw, Poland\\
$^{21}$Aix Marseille Univ, CNRS, CNES, LAM, Marseille, France\\
$^{22}$INAF, Istituto di Radioastronomia, Via Gobetti 101, I-40129 Bologna, Italy\\
$^{23}$Department of Physics and Astronomy, University of the Western Cape, Robert Sobukwe Road, Bellville 7535, South Africa\\
$^{24}$Sterrenkundig Observatorium, Universiteit Gent, Krijgslaan 281 S9, B-9000 Gent, Belgium\\
$^{25}$International Centre for Radio Astronomy Research (ICRAR), University of Western Australia, 35 Stirling Highway, Perth, WA 6009\\
$^{26}$ASTRON, the Netherlands Institute for Radio Astronomy, Postbus 2, 7990 AA, Dwingeloo, The Netherlands\\
$^{27}$Research Center for Intelligent Computing Platforms, Zhejiang Laboratory, Hangzhou 311100, China\\
}
\date{Accepted 2023 March 15. Received 2023 March 15; in original form 2022 November 11}

\pubyear{2022}

\begin{document}
\label{firstpage}
\pagerange{\pageref{firstpage}--\pageref{lastpage}}
\maketitle
\begin{abstract}
The transformation and evolution of a galaxy is strongly influenced by interactions with its environment. Neutral hydrogen (\HI) is an excellent way to trace these interactions. Here, we present \HI\ observations of the spiral galaxy NGC~895, which was previously thought to be isolated. High-sensitivity \HI\ observations from the MeerKAT large survey project MIGHTEE reveal possible interaction features, such as extended spiral arms, and the two newly discovered \HI\ companions, that drive us to change the narrative that it is an isolated galaxy. We combine these observations with deep optical images from the Hyper Suprime Camera to show an absence of tidal debris between NGC 895 and its companions. We do find an excess of light in the outer parts of the  companion galaxy MGTH$\_$J022138.1-052631 which could be an indication of external perturbation and thus possible sign of interactions. Our analysis shows that NGC~895 is an actively star-forming galaxy with a SFR of $\mathrm{1.75 \pm 0.09 [M_{\odot}/yr]}$, a value typical for high stellar mass galaxies on the star forming main sequence. It is reasonable to state that different mechanisms may have contributed to the observed features in NGC~895 and this emphasizes the need to revisit the target with more detailed observations. Our work shows the high potential and synergy of using state-of-the-art data in both \HI\ and optical to reveal a more complete picture of galaxy environments.
\end{abstract}



\section{Introduction}
Numerous observations have shown that galaxy properties such as morphology, size, star formation and kinematics are strongly influenced by their connections to the surrounding environment (\citealp{1980ApJ...236..351D,2010ApJ...721..193P,2017MNRAS.465.4572Z}). Most galaxies in the nearby universe are either interacting with or gravitationally bound to nearby companions (e.g, \citealt{Okamoto_1999,2017MNRAS.465.4572Z,2009ARA&A..47..159B}). Such interactions are directly responsible for an abrupt change in galaxy properties. According to the prevailing theoretical model for structure formation, galaxies are surrounded by a halo of dwarf galaxies and these galaxies form and evolve hierarchically through a succession of mergers \citep{1978MNRAS.183..341W}. For example, the most massive local group satellite, the Large Magellanic Cloud (LMC), is one of many dwarf galaxies that surround the Milky Way galaxy \citep{2010AdAst2010E..21W,2015ApJ...805..130K}. While interaction between massive galaxies can cause powerful star formation events \citep{1990ApJS...74..833H}, interactions between dwarf companions and massive galaxies can also be dramatic (e.g., the Magellanic Stream being pulled from the SMC/LMC system by interactions with the Milky Way; \citealt{1998Natur.394..752P}).

Close encounters of galaxies can show significant signatures of gravitational interaction in the form of a distortion in their structure, the presence of tails and bridges, or a common diffuse envelope (\citealp{1992ARA&A..30..705B,2013ApJ...779L..15N}). All these features have been quantitatively explained based on numerous N-body simulations, starting with those of \citet{1972ApJ...178..623T}. The $\Lambda$-CDM cosmology assumes that apart from dark haloes with normal (luminous) galaxies in their centres, completely dark clumps (sub-haloes) should also exist with masses $\sim$ 10$^{8}$ - 10$^{11}$ M$_{\odot}$ \citep{2005MNRAS.358..217Y}. If completely dark galaxies exist \citep[e.g., ][]{2021MNRAS.507.2905W}, then the evidence of interactions may also occur in the case of galaxies that appear isolated \citep[but see][]{2021A&A...649L..14R}. Apart from interaction with a dark galaxy, irregularities of isolated galaxies may be due to an earlier merger with a now consumed companion \citep{2006A&A...451..817K}, or a brief, fly-by interaction with a companion currently far away \citep{2012ApJ...751...17S}. There could additionally be large gas accretion from cosmic filaments \citep{2005MNRAS.363....2K} and asymmetric accretion that could also lead to perturbed morphologies. 

One way to identify interacting/merging galaxies is through the characterization of their optical morphological properties. Late-stage mergers are thus identified as having highly disturbed morphologies \citep{2019MNRAS.485.5631C}, while early-stage merging galaxies are often characterized by the presence of tidal features. Close pairs of galaxies (merger candidates) on the other hand mark the beginning stages of the merger process. The states of non-equilibrium induced by merger activity in both the stellar and dark matter components of a galaxy translate into asymmetries in the galaxy stellar light distribution \citep{2009ApJ...691.1005R}, and it is these asymmetries that have become useful tracers of merger activity in the optical regime.

Neutral hydrogen gas (\HI), however, remains one of the best tracers for galaxy interactions because the \HI\ envelope of a galaxy is generally more extended than its optical disk \citep{1996ApJS..105..269H, Michel_Dansac_2010,2018MNRAS.478.1611K}, and is more easily disrupted by tidal forces from neighbouring galaxies. Galaxy pairs often show complex extended \HI\ distributions, including tails and bridges \citep{2004AJ....128...16K}, even if the corresponding optical images show few disturbances. Observations of the \HI\ distribution and kinematics of large numbers of spiral galaxies now provide abundantly detailed information on the nature of interactions/mergers in both the \HI\ disc kinematics and the \HI\ distribution, which influences the shape of the global profile \citep{1999MNRAS.304..330S,2020MNRAS.493.2618S}. \citet{2012ApJ...760L..25E} quantified the range \HI\ profile asymmetries in a sample of some of the most isolated galaxies in the local Universe (AMIGA, \citealt{2005A&A...436..443V}), and found that it shows the lowest level of HI-asymmetry of any galaxy sample. Galaxies with asymmetric global profiles often have rotation curves that rise more slowly on one side of the galaxy than on the other \citep{2008A&ARv..15..189S, 1999MNRAS.304..330S,2020MNRAS.495.1984D}. Correspondingly, iso-velocity contours of the velocity field show a steeper gradient. All this may indicate past or present interaction activity.

High-sensitivity \HI\ observations have typically been conducted in the past with single-dish telescopes (e.g, \citealt{2001MNRAS.322..486B, Giovanelli_2005, 2006AJ....131.2913B}) which are ideal for probing low surface brightness \HI\ emission, as these instruments do not filter out emission on any scales \citep{2014AJ....147...48P}. However, single dishes have coarse spatial resolution that is often unable to resolve the \HI\ in nearby galaxies, especially companion dwarf galaxies which tend to be fainter and smaller. The Square Kilometre Array (SKA) pathfinder and precursor telescopes such as MeerKAT \citep{2018ApJ...856..180C} have been built with a unique combination of high sensitivity, high angular resolution and wide field of view which allows us to survey the sky at unprecedented speeds and improved sensitivity.

The MeerKAT International GHz Tiered Extragalactic Exploration \citep[MIGHTEE;][]{2016mks..confE...6J} Survey is a medium-deep, medium-wide blind survey carried out with MeerKAT. The Early Science MIGHTEE observations provide \HI\ line cubes with relatively high sensitivity to diffuse and faint emission. MIGHTEE \HI\ observations can thus help to understand the behaviour of \HI\ gas-rich galaxies and their surrounding environments (e.g. \citealt{2021MNRAS.506.2753R}). In this paper, we are particularly interested in one of the galaxies contained within the survey volume, NGC~895. It is an exciting system to study since this nearby galaxy, classified as a SA(s)cd \citep{1991rc3..book.....D} has been classified as isolated \citep{1988ngc..book.....T}, but in fact, it exhibits strong signs of interaction in \HI\ such as extended tidal arms and a warp. \citet{2002ApJS..142..161P}, using Very Large Array (VLA) data, report an \HI\ mass of 1.3 $\times$ 10$^{10}$ M$_{\odot}$ and reveal a severely warped inclined galaxy, although no obvious companion was observed. The star formation properties of NGC~895 were derived by \citet{2021A&A...646A..35M} using multi-wavelength optical and near-infrared imaging to perform photometry. Using the HSC \textit{G}-band, GFHT \textit{u}-band, HSC \textit{riZY}, and near-infrared \textit{JHKs} bands, they fit the photometry using spectral energy distribution (SED) fitting to derive a stellar mass of 7.86\,$\times$ 10$^{9}$ M$_{\odot}$, a star formation rate of 7.3\,$\times$ 10$^{-10}$ M$_{\odot}$/yr and an age of 2.5\,Gyr. Note that the values quoted by \citet{2021A&A...646A..35M} were all computed in bulk, i.e. objects were not treated individually. Some basic information on NGC~895 can be found in Table~\ref{table1}.

In this paper, we use high sensitivity, high spatial resolution MeerKAT observations as part of the MIGHTEE-\HI\ early science, complemented by Subaru deep optical images to explore the environment surrounding NGC~895. We aim to search for potential companions and determine their effects on the properties of NGC~895. In section \ref{sec:obs} we discuss the observations and data reduction techniques of our \HI\ observations and also present the auxiliary optical data. In section \ref{sec:hicontent} we present the \HI\ distribution. Section \ref{sec:rotation} provides the analysis of the \HI\ kinematics of NGC~895, and Section \ref{sec:star} presents the star formation properties of NGC~895. In Section \ref{sec:optical} we provide the optical analysis of two newly detected \HI\ companions, while Section \ref{sec:discuss} provides a detailed discussion of the possible causes of the observed properties in NGC~895. Lastly, the conclusions are given in Section \ref{sec:conclude}. 

Throughout this paper, we assume $\Lambda$CDM cosmology parameters of $\Omega_{m}$ = 0.315, $\Omega_{\Lambda}$ = 0.685, $H_{0}$ = 67.4 \kms Mpc$^{-1}$ and AB magnitudes are used throughout unless otherwise stated. The distances used in this paper are cosmological luminosity distances, calculated using \HI\ redshifts and assuming the cosmological values stated above.
\begin{table}
\caption{\small Basic Properties of NGC~895}
\begin{minipage}{\textwidth}
\begin{tabular}{l@{\hspace{0.70cm}}c@{\hspace{0.05cm}}}   
\hline   
Parameter &NGC 895 \\
                            
\hline \hline  
Morphology  &SA(s)cd$^{(a)}$	\\
Right ascension (J2000) &02:21:36.5 \\
Declination (J2000)&-05:31:17.0 \\
Distance (Mpc) & 34.3 $\pm$ 1.1 \\
V$_{\text{heliocentric}}$ (km.s$^{-1}$) &  2286.0 $\pm$ 22\\
PA$\_$opt($^\circ$) &110.0 \\
Inclination$\_$opt($^\circ$) & 39.0\\
Redshift & 0.007658 $\pm$ 0.00006 \\
Total \HI\ mass (M$_{\odot}$)&1.3 $\times10^{10}$ $^{(b)}$\\
Stellar mass (M$_{\odot}$) & (2.28 $\pm$ 0.16) $\times 10^{10}$ $^{(c)}$ \\
SFR (M$_{\odot}$/yr) & 1.75 $\pm$ 0.07 $^{(c)}$\\
Age (Gyr) & 2.50\\
\hline     \\
\multicolumn{2}{@{} p{8 cm} @{}}{\footnotesize{\textbf{Notes.} Ref\,(a) \citet{1991rc3..book.....D}; (b) \citet{2002ApJS..142..161P}; (c) this work. All other values are derived from the MIGHTEE-\HI\ data using the method described in \citet{2021A&A...646A..35M}}.}
\label{table1}
\end{tabular}   
\end{minipage}
\end{table} 
\section{Observations}\label{sec:obs}
\subsection{The MIGHTEE Survey -- Observations and Data Reduction} 
MIGHTEE targets four well-known deep extragalactic fields. The data are collected in full spectral line and polarization modes. The \HI\ emission project within the MIGHTEE survey is referred to as MIGHTEE-\HI, described in detail in \citet{2021A&A...646A..35M}. This work uses the MIGHTEE-\HI\ Early Science observations, which were conducted between mid-2018 and mid-2019. These observations were performed with the full array (64 dishes, minimum 58 dishes) in the L-band, which has a frequency range from 900 to 1670 MHz. We used the MeerKAT 4k correlator mode, which has 4096 channels, and a channel width of 209 kHz, corresponding to 44 \kms at \textit{z} = 0. 

The data reduction was performed with the \textsc{Process-MeerKAT} calibration and imaging pipeline (Frank et al. in prep.). The pipeline is \textsc{casa} - based \citep{2007ASPC..376..127M} and does spectral line calibration following the standard data reduction and calibration tasks such as flagging, delay, bandpass and complex gain calibration. Spectral-line imaging was performed using \textsc{casa} TCLEAN with the Briggs (\textsc{ROBUST} = 0.5) weighting scheme. The continuum subtraction was done in the visibilities using standard \textsc{casa} routines UVSUB and UVCONTSUB. Once the cubes were produced, per-pixel median filtering was applied to the resulting data cubes to reduce the impact of the direction-dependent artefacts. The restored synthesized beam for the \HI\ cube has a size of $12^{\prime \prime} \times\ 9^{\prime \prime}$ with a position angle of -2.6 deg. The \HI\ cube has a rms noise per 44 \kms channel of 85 $\mu$ Jy\ beam$^{-1}$, which translates to a 3$\sigma$ column density sensitivity of N$_{\text{\HI}}$ = \num{1.2e+20} $\mathrm{cm^{-2}}$. 

Moments maps were constructed using the {\tt{spectral-cube}} software\footnote{https://spectral-cube.readthedocs.io/en/latest/} and the standard procedure as described in \citet{2021MNRAS.508.1195P}. We created a cubelet from the MIGHTEE \HI\ field centred on NGC~895. The cubelet was then smoothed to a circular beam of 20$^{\prime \prime}$ and clipped at the 3$\sigma$ level. The resulting mask was then applied to the original resolution cubelet to account for the low column density extended \HI\ emission. Lastly, the remaining noise pixels were masked out by hand. 
\begin{table}
\caption{Summary of MeerKAT MIGHTEE-\HI\ data products used in this work \citep{2021A&A...646A..35M}}
\begin{tabular}{cc}  
\hline \hline  
Number of antennas & 58 to 64 \\
Total integration & 13 h on source \\
FWHM of primary beam & $\sim$1 $^{\circ}$\\
Channel width & 209 kHz \\
Number of channels & 4096 \\
\hline  
\multicolumn{2}{c}{Robust = 0.5 weighting function} \\ \\
FWHM of synthesized beam & $12^{\prime \prime} \times\ 9^{\prime \prime}$ (2.0 kpc $\times$ 1.5 kpc)  \\
RMS noise per channel & 85 $\mu$ Jy\\
\HI\ column density limit & \\
(3$\sigma$ over 1 channel, $12^{\prime \prime} \times\ 9^{\prime \prime}$) & $\sim 1.2 \times\ 10^{20}$ cm$^{-2}$ \\
(3$\sigma$ over 1 channel, $20^{\prime \prime} \times\ 20^{\prime \prime}$) & $\sim 3.1 \times\ 10^{19}$ cm$^{-2}$ \\
\hline    
\end{tabular}
\label{coords_table}
\end{table}  
\subsection{Optical data}
We use the deepest optical dataset that overlaps with MeerKAT images in this region of the sky, from the Hyper Suprime-Cam Subaru Strategic Program Survey (HSC-SSP) \citet{2018PASJ...70S...4A}. This survey makes use of the Subaru 8.2 m telescope providing images with sub-arcsec resolution in the broad bands \textit{g, r, i, z} and \textit{y}. We downloaded mosaics for each of the bands, including a wide enough field around NGC~895 using the \texttt{hscmap} browser hosted on the survey web page: \url{https://hsc-release.mtk.nao.ac.jp/doc/}. We used the second data release of the survey \citep{2019PASJ...71..114A}. We calculated the limiting surface brightness of these mosaics following the technique applied by \cite{2020A&A...644A..42R}, their appendix A. The 3$\sigma$ values in $10^{\prime \prime} \times\ 10^{\prime \prime}$ boxes are 30.2, 29.6, 29.4, 28.7 and 27.7 mag arcsec$^{-2}$ in \textit{g, r, i, z} and \textit{y}, respectively. The 5$\sigma$ point source detection limits are \textit{g}~=~26.8, \textit{r}~=~26.4, \textit{i}~=~26.2, \textit{z}~=~25.4, \textit{y}~=~24.7 mag according to the survey.

Additionally, we use data from the Dark Energy Camera Legacy Survey \citep[DECaLS;][]{2019AJ....157..168D}. This is a multi-purpose deep optical survey that makes use of the DECAM imager on the 4\,m Blanco Telescope in the three optical bands \textit{g, r} and \textit{z} with a seeing of the order of 1$^{\prime \prime}$. The depths measured in the region of NGC~895 are 29.3, 29.0 and 27.8 for the \textit{g, r} and \textit{z} bands at 3$\sigma$; $10^{\prime \prime} \times\ 10^{\prime \prime}$  boxes. The 5$\sigma$ point source detection limits are \textit{g}~=~24.7, \textit{r}~=~23.6 and \textit{z}~=~22.8 mag according to the survey. Although the DECaLS data is less deep than the previously mentioned HSC-SSP data, the DECaLS data has a processing step in which the stars are subtracted by modelling the PSF, alleviating problems with confusion and foreground Galactic stars, which is relevant for Section \ref{sec:optical}.
\section{NGC~895 analysis}
\subsection{\HI\ content and distribution} \label{sec:hicontent}
\begin{figure*}
\centering
   \advance\leftskip0cm
   \includegraphics[height=8.5cm]{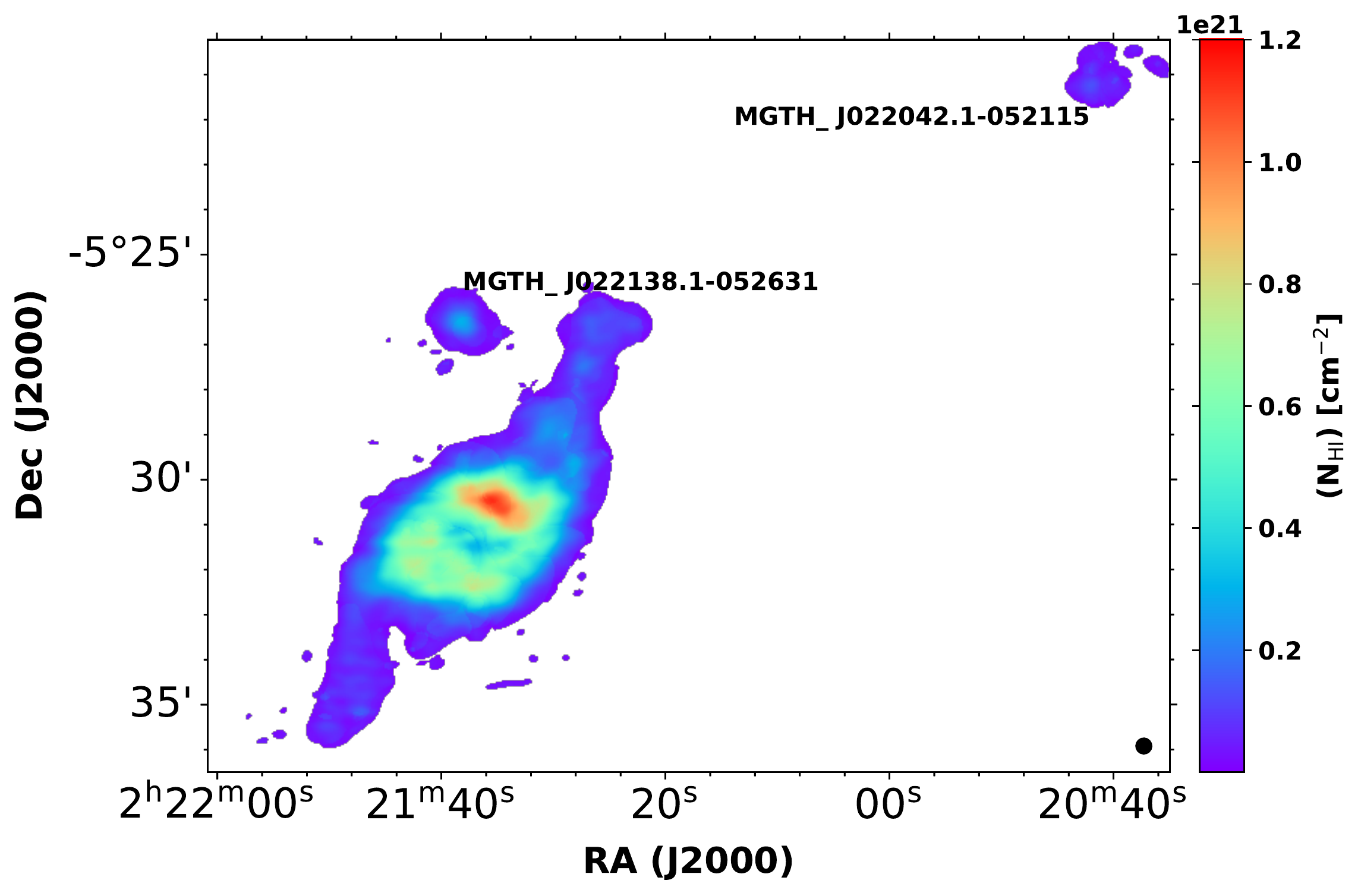}

   \includegraphics[height=10cm]{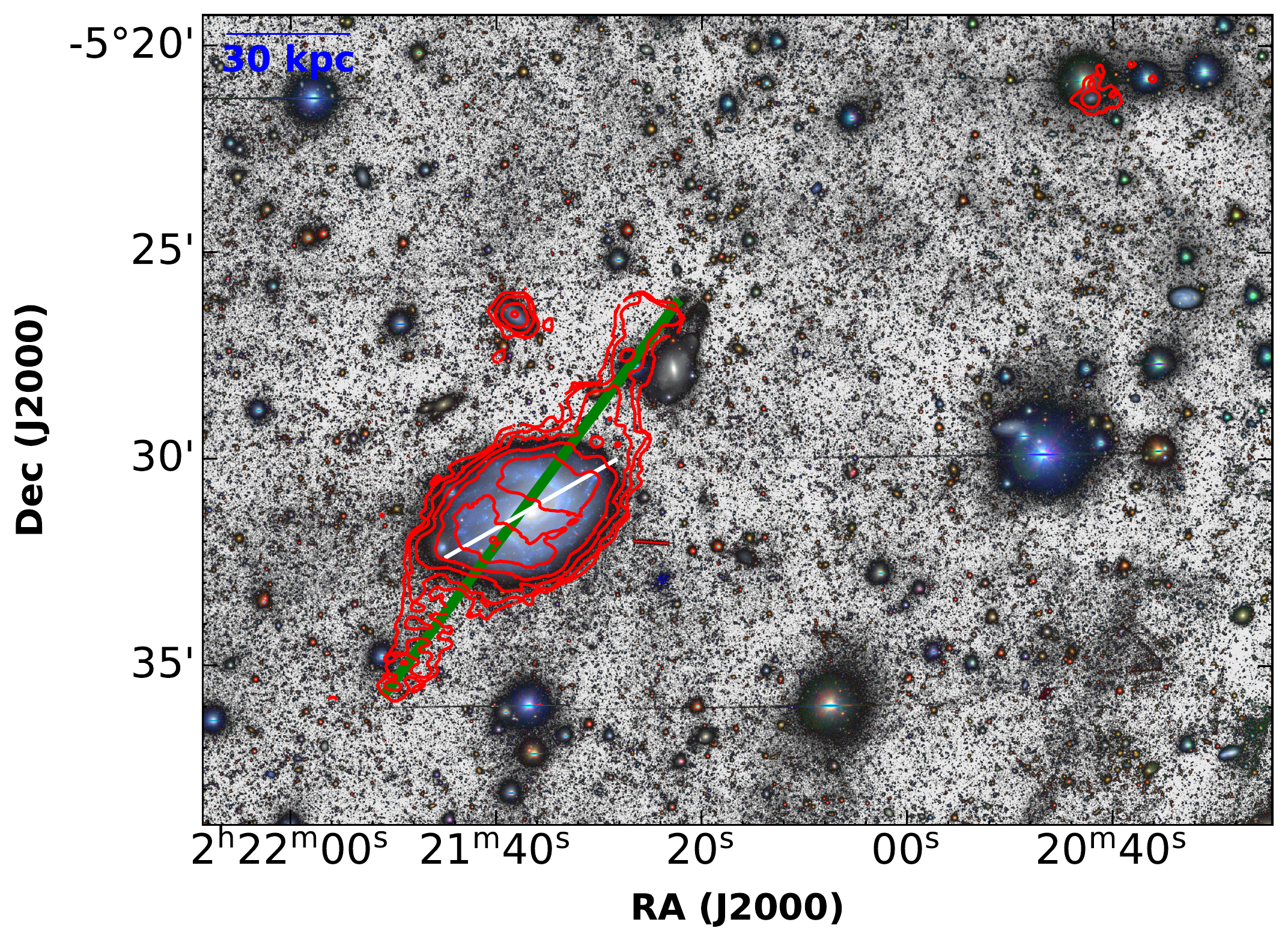}
   \advance\leftskip-1.8cm
\caption{\textbf{Top:} Moment 0 map of the \HI\ emission in NGC~895. The black circle in the bottom right corner indicates the beam size (20$^{\prime \prime} \times 20^{\prime \prime}$). The newly detected \HI\ companions of NGC~895 are labelled. At the galaxy centre, NGC~895 shows a depression in the \HI\ column density. \textbf{Bottom:} MeerKAT H\,{\sc i} column density map of NGC~895 overlaid on a deep HSC 3-band composite image. Contour levels are (3.2 $\times$ 10$^{19}$, 6.4 $\times$ 10$^{19}$, 1.28 $\times$ 10$^{20}$, 2.56 $\times$ 10$^{20}$, and 5.12 $\times$ 10$^{20}$) atoms\ cm$^{-2}$. The white line represents the estimated optical diameter of NGC~895 derived from the g-band color image, while the green line shows the estimated \HI\ diameter of NGC~895.}
\label{fig:mom0}
\end{figure*}

The \HI\ moment-0 map of NGC~895 and companion galaxies is shown in thevtopvpanel of Fig.~\ref{fig:mom0}, and the \HI\ column density map, overlaid on a deep HSC 3-colour image, is shown in the bottom panel. The sensitivity of our observations allows us to detect two \HI\ companions located to the north and north-west of the main galaxy (see Fig. 1), which we identify as MGTH$\_$J022138.1-052631 and MGTH$\_$J022042.1-052115 (referred as J022138 and J022042 hereinafter). The two companions have spatially coincident optical emission, and are at projected distances of 46.5 and 173.7\,kpc from NGC~895, respectively. The inner region of NGC~895 appears to have a normal morphology typical for spiral galaxies \citep{2013seg..book..155B}. The lowest \HI\ contour corresponds to the 3.2$\sigma$ column density limit of \num{3e+19} $\mathrm{cm^{-2}}$. At this level, the galaxy has a diameter of 115\,kpc at our adopted distance of 34.3 $\pm$ 1.1\,Mpc. This is $\sim$ 2.7 times the optical diameter of 43 kpc derived from the g-band color image. The \HI\ gas distribution, as seen in Fig.~\ref{fig:mom0} reveals an \HI-poor central region.

\begin{figure}
\centering
   \includegraphics[height=5cm]{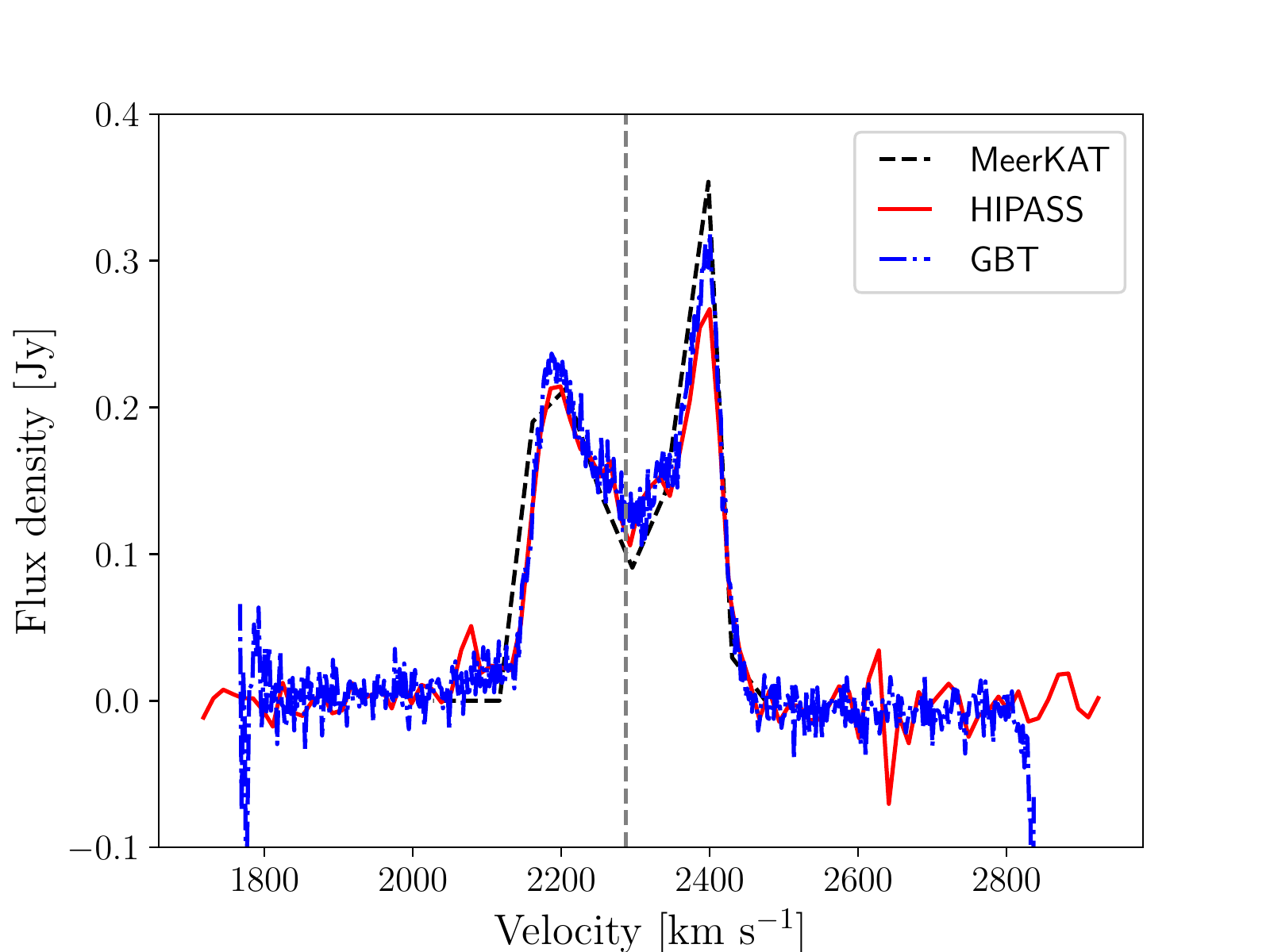}
\caption{H\,{\sc i} spectral profile from our MeerKAT observations (black dash-dotted line), from the GBT \citep{2005ApJS..160..149S} (blue dash-dotted line) and Parkes\citep{2004AJ....128...16K} (red line). The grey vertical line shows the \HI\ systemic velocity of NGC~895 (V$_{sys}$ = 2286 \kms) derived from our kinematics.}
\label{fig:prof}
\end{figure}

Figure~\ref{fig:prof} shows the global \HI\ profile of NGC~895. It has been derived from the primary-beam corrected cubes. We compare the global profile of NGC~895 as obtained from MeerKAT, GBT and HIPASS observations. It is clear from Fig.~\ref{fig:prof} and Fig.~\ref{fig:rot} that NGC~895 is asymmetric, with more \HI\ on the receding than on the approaching side.
The \HI\ line flux derived from the global profile of NGC~895 as measured from the MeerKAT data is 52.1 $\pm$ 0.3\,Jy.\kms. This is in agreement with the GBT and HIPASS fluxes of 50.51\,Jy.\kms \citep{2005ApJS..160..149S} and 49.4\,Jy.\kms \citep{2004AJ....128...16K}, respectively. Using the \HI\ mass standard formula, 
\begin{equation}
 M_{\text{HI}} = 2.36 \times 10^{5} D^{2} \int F d\nu \times \frac{1}{1 + z}
\end{equation}
where $D$ is the distance in Mpc, $F$ is the flux density in Jy, d$\nu$ is the velocity resolution in \kms and $z$ is the redshift, the corresponding total \HI\ mass is 1.4 $\pm$ 0.01 $\times$ 10$^{10}$ M$_{\odot}$. This value is $\sim$ 8$\%$ higher than that of \citet{2002ApJS..142..161P}. 

\subsection{\HI\ Kinematics}\label{sec:rotation}
\begin{figure}
\centering
   \includegraphics[height=5cm]{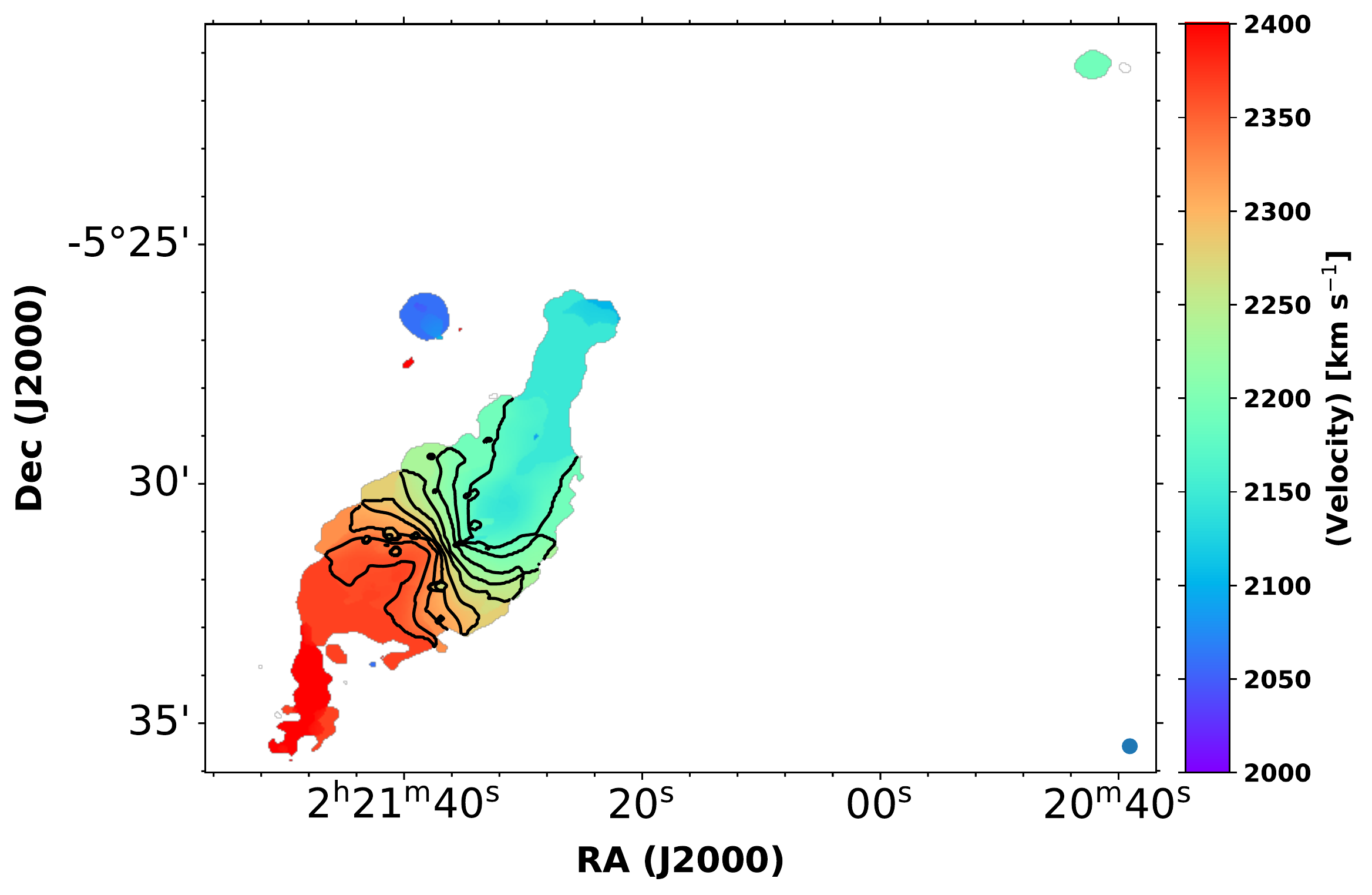}
\caption{The MeerKAT H\,{\sc i} line of sight velocity of NGC~895. The velocity contours are 2180, 2200, 2220, 2240, 2260, 2280, 2300, 2320, 2340, and 2360 \kms.}
\label{fig:rot}
\end{figure}

The velocity field map of NGC~895 is shown in Fig. ~\ref{fig:rot}. The map shows a different curvature of the iso-velocity contours on the approaching and receding side, which suggests the presence of kinematic asymmetry \citep{2021A&A...654A...7K}. We use an extended version of the tilted-ring model by \citet{1974ApJ...193..309R} implemented in the TIlted RIng FIitting Code {\tt {TiRiFiC}} \citep{2007A&A...468..731J} to model the kinematics and the gas distribution in NGC~895. {\tt{TiRiFiC}} uses a $\chi^{2}$ minimization technique to fit a tilted-ring model directly to a data cube instead of a velocity field. This makes {\tt{TiRiFiC}} less prone to beam smearing effects and represents more accurately complex kinematical features, such as a warp, compared to the classic method of fitting to a velocity field. {\tt{TiRiFiC}} models a rotating gas disc as a set of concentric annuli, each described by the user-supplied input parameters such as the central coordinates (XPOS, YPOS), the rotation curve (VROT), the position angle (PA), the inclination (INCL), the thickness of the disk (Z0), and potentially many more. 

To make a reasonable initial guess for the {\tt {TiRiFiC}} model parameters, we use the Fully Automated {\tt {TiRiFiC}} (FAT) software \cite{2015MNRAS.452.3139K}, which is a  GDL wrapper around {\tt {TiRiFiC}}. We used the beta version of FAT, which allows for radial variation of the intrinsic velocity dispersion. FAT takes an \HI\ data cube as input and automatically estimates the following free parameters; 1) surface brightness, 2) position angle, 3) inclination 4) rotational velocity, 5) systemic velocity, 6) scale height 7), intrinsic dispersion and 8) central coordinates. FAT fits each parameter ring by ring and finally smooths the parameters with a polynomial of order 0, 1, 2, 3, 4 or 5. FAT models the \HI\ disc as two halves and fits rings across each half. The procedure employed by FAT for finding the initial estimates of the free parameters and the flow-chart for fitting is discussed in detail in \citet{2015MNRAS.452.3139K}. The results from FAT were then inputted to {\tt {TiRiFiC}}, allowing us to adjust the initial estimates manually and, if necessary, add parameters not implemented by FAT. 

Unlike {\tt {TiRiFiC}}, FAT is fully automated and no parameter tweaks are required to optimize the results. The output from FAT is a very good starting point for {\tt {TiRiFiC}}. {\tt {TiRiFiC}} is only necessary if we want to fit non-axisymmetric features not captured by FAT, or if we want to fit the galaxies as one disk (i.e., not as a receding and approaching halves). The stopping criteria for TiRiFiC are reached by visual inspection of various plots of the model and the observed data cube (Position-velocity diagrams, residual velocity field, moment maps, and channel maps). Based on this, we judge whether certain model features are required to better reproduce the observations or not. The degeneracy between velocity and inclination, which is often seen for 2D fitting, is broken for {\tt {TiRiFiC}} if the disk is symmetric and has a gradient in surface brightness. For the case of a very thick disk, the inclination, rotation velocity, and disk thickness are degenerate. For NGC 895, the disk thickness is not high enough to cause such a degeneracy.

To minimize the number of free parameters, we fit the galaxy as one disc instead of modelling the approaching and the receding halves separately, as done by FAT. However, to capture the asymmetric distribution in surface brightness, we introduce a first order azimuthal variation in amplitude (SM1A) and phase (SM1P). Both of these parameters were varied  with radius. Errors were calculated using the bootstrap method described in \citep{2020MNRAS.497.4795I, Jozsa2021}. The method consists of first getting model parameters using a Golden-Section nested intervals algorithm. The model parameter at a single node is  then shifted by a random value to create a synthetic datum. After creating many such data and fitting each of them, the mean value and the standard deviation of the best-fitting parameters are taken as the final model parameters and the standard errors, respectively.     

\begin{figure*}
\centering
   \includegraphics[scale=0.46]{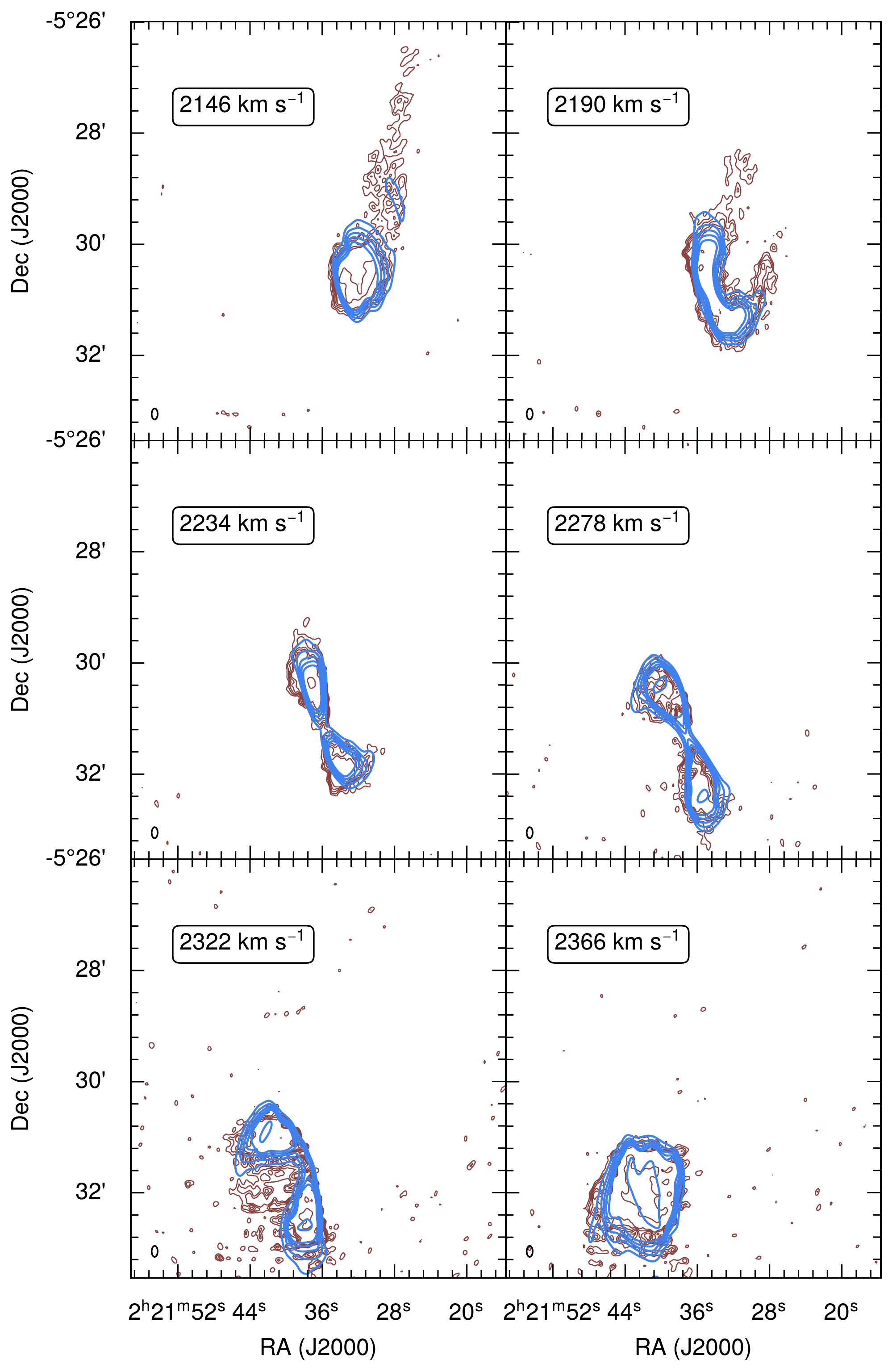}
    \includegraphics[scale=0.361]{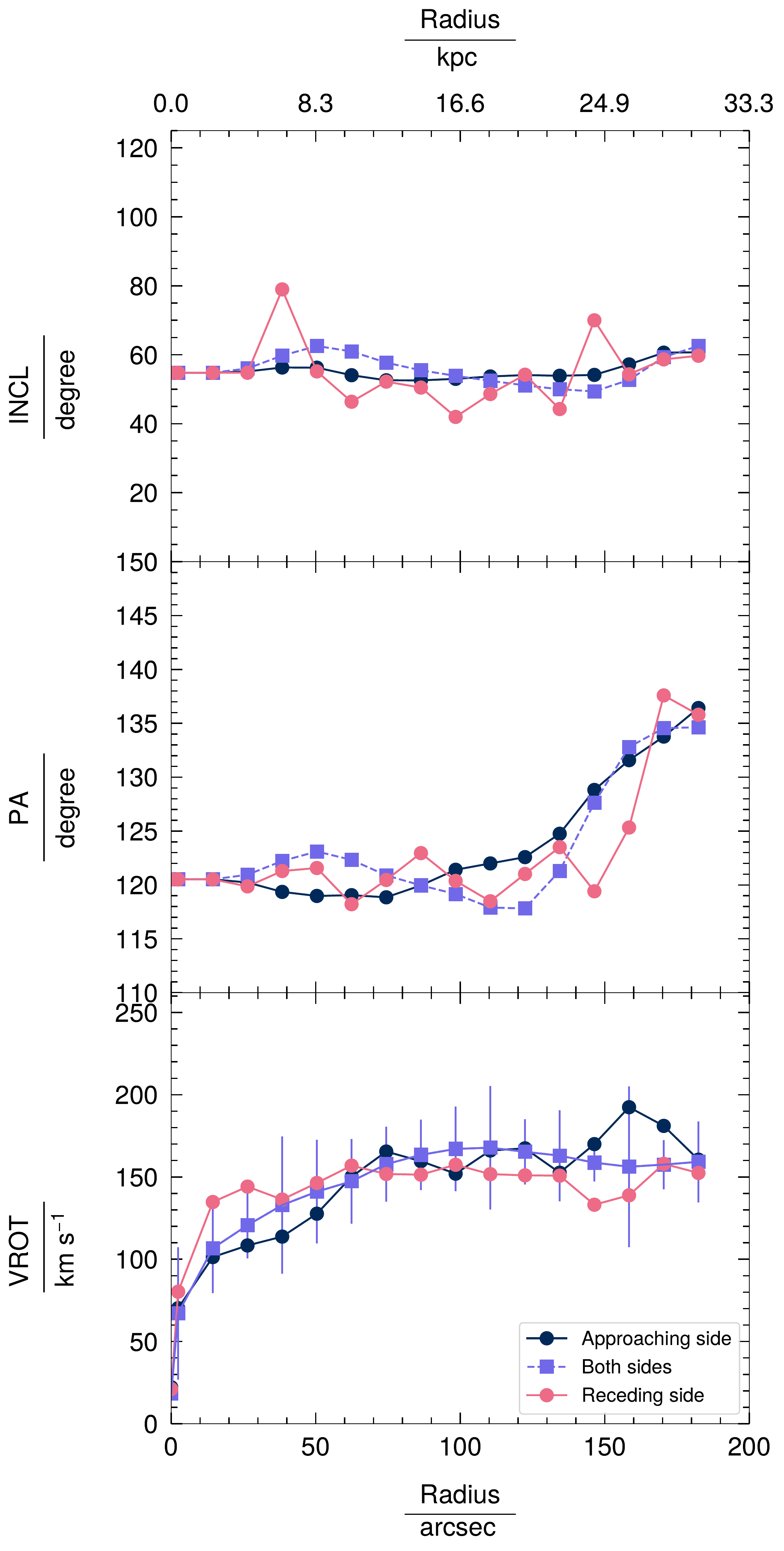}
\caption{Left: MeerKAT \HI\ channel maps of NGC~895 (brown contours) in comparison with the corresponding channel maps of the {\tt {TiRiFiC}} model (blue contours) in the velocity range between 2102 and 2410 \kms. The velocity of each channel is shown in the top left. Contours are at (0.78, 1.17, 1.56, 1.95, 2.34, 4.69) mJy/beam. Right: {\tt {TiRiFiC}} best-fit solutions for the inclination (top panel), the position angle (middle panel) and the rotation curve (bottom panel).}
\label{fig:chan}
\end{figure*}

Figure~\ref{fig:chan} (left) compares the data (brown) and the model (blue) at different velocities. The main disc of the galaxy is well modelled and shows a good resemblance to the data. The presence of \HI\ emission not captured by our model is also seen in Fig. ~\ref{fig:chan} (left). These indicate asymmetric features (arms) most likely induced by an interaction. In Fig.~\ref{fig:chan} (left), there are extensions in the data (e.g. channel 2146 \kms) clearly pointing North, in the direction of the companion J022138. This suggests the present gravitational interaction between those two objects. As we will see in the next section, this is also the companion showing an excess of light in the outer parts, another indication of possible interaction in the past. Fig.~\ref{fig:chan} (right) shows {\tt {TiRiFiC}} results for the inclination, position angle and rotation velocity of NGC~895. At a radius of $\sim$ 16 kpc, we see a major shift in the position angle, which is interpreted as being due to a warp.

\begin{figure*}
\centering
   \begin{tabular}{cc}
   \includegraphics[scale=0.65]{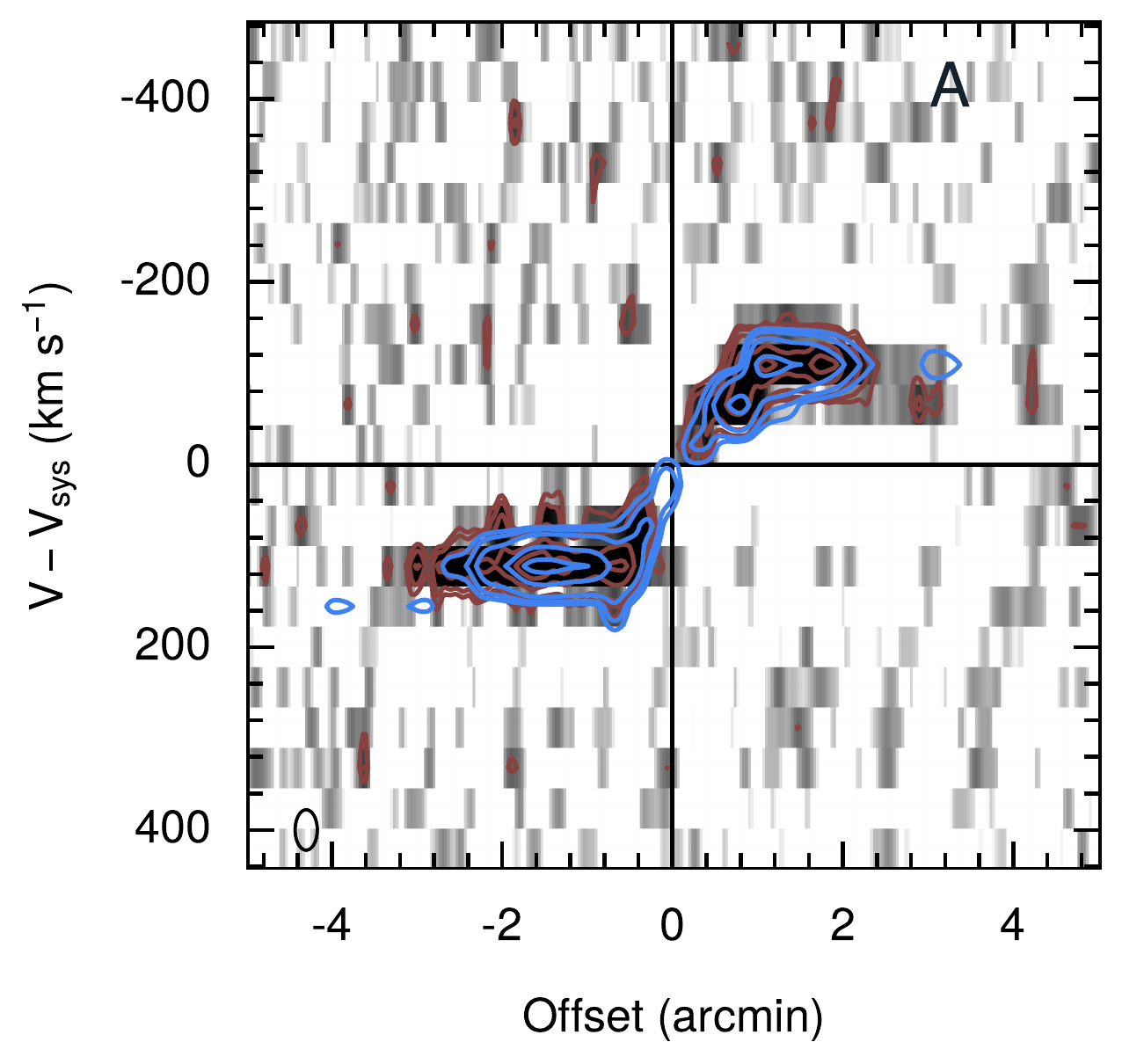} & 
    \includegraphics[scale=0.65]{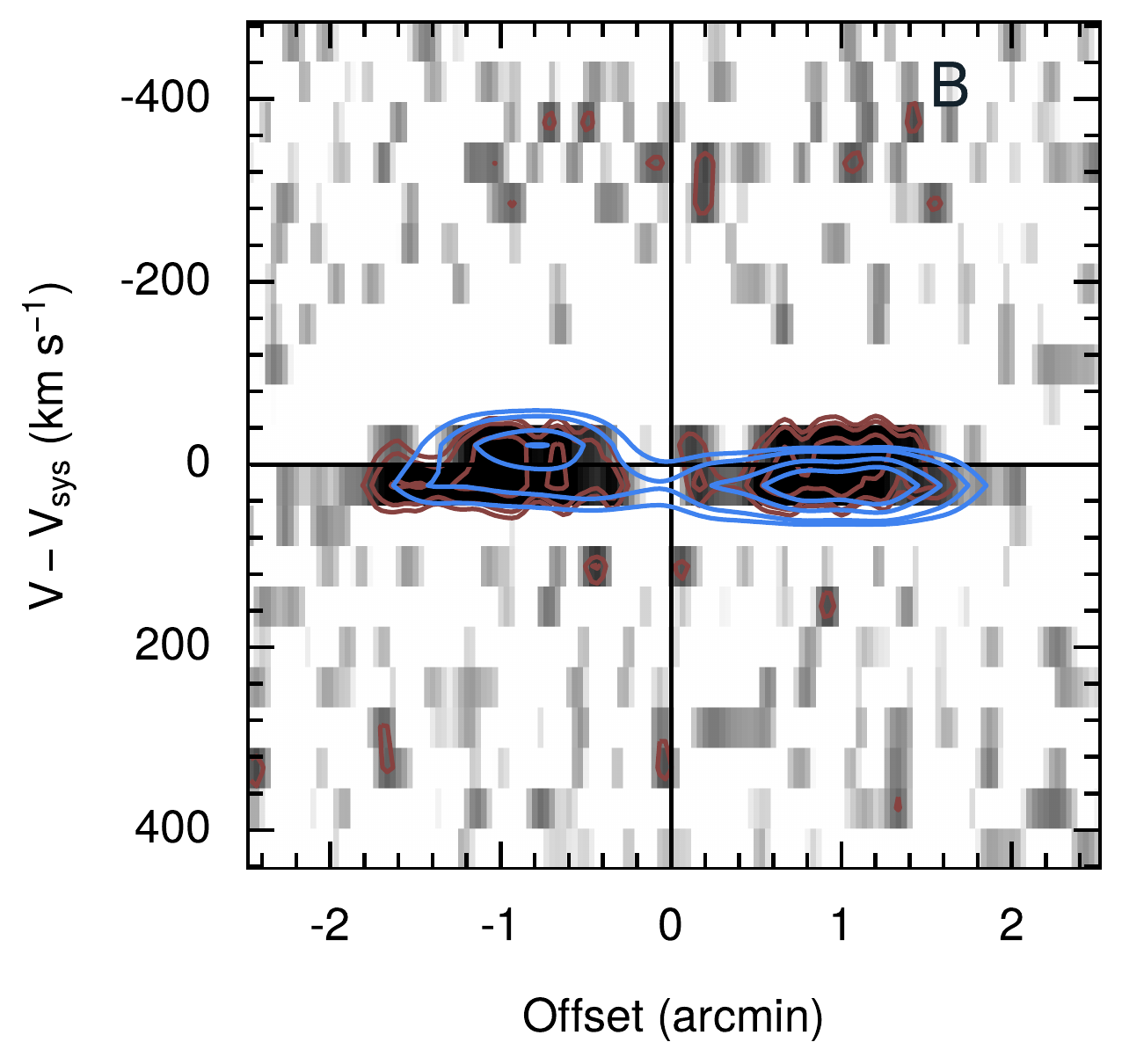}
   \end{tabular}
\caption{Comparison between the position-velocity diagrams of the observed data cube (brown contours) and the TiRiFiC model data cube (blue contours). The contour levels are (0.6, 0.9, 1.5, 3. , 4.5) $\mathrm{mJy~beam^{-1}}$. Panel A was taken from a slice at \ang{120} (the inner disk's position angle) along the major axis. Panel B was taken from a slice perpendicular to slice A at the kinematic centre.}
\label{fig:model-obs}
\end{figure*}

Figure~\ref{fig:model-obs} shows the position-velocity diagram of NGC~895 along the major axis and minor axis, with the data (brown) overlaid on the model (blue). We see that the model is consistent with the data. From the position-velocity diagram, we see no traces of extraplanar gas, which suggest that the arms we see in NGC~895 are asymmetric features induced by an interaction. The systemic velocity derived for NGC~895 is 2253 $\pm$ 32 \kms. This value is consistent with the value of 2286 $\pm$ 20 \kms derived from the \HI\  global profile. The rotation curve is derived out to a radius of R$_{\textsc{HI}}$ = 184 $^{\prime \prime}$ (30.5 kpc), where the rotation velocity reaches a value of 159 $\pm$ 24 \kms.

\subsection{Star formation \& stellar mass estimates}\label{sec:star}

\begin{figure}
    \centering
    \includegraphics[width=8.5cm]{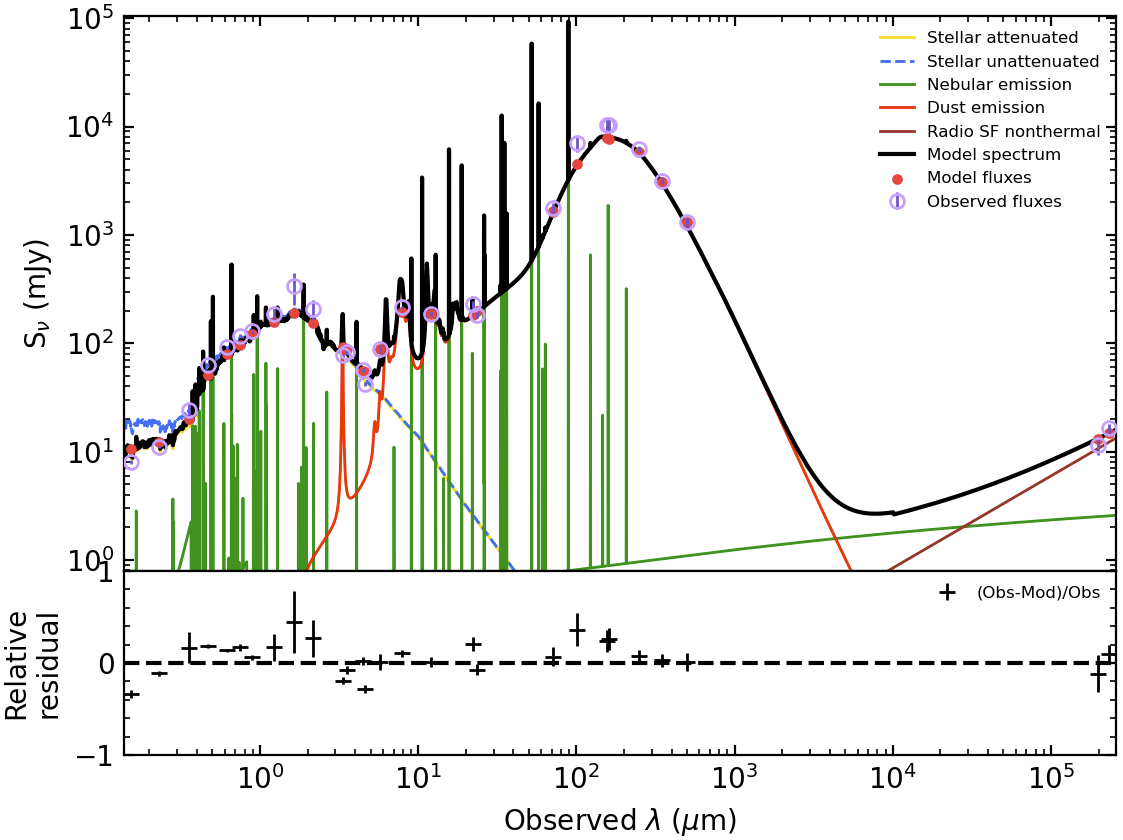}
    \caption{\texttt{CIGALE} SED Fitting results of NGC~895. Best fit model parameters are reported in Sec.~\ref{sec:star}. Observed fluxes are plotted with open  violet circles. Filled red circles correspond to the model fluxes. The final best model is plotted as a solid black line. Radio emission is shown as a brow line, while the remaining four lines correspond to the stellar, dust, and nebular components. The relative residual fluxes, calculated as (observed flux– best model flux)/observed flux, are plotted at the bottom of the panel.}
    \label{fig:SED}
\end{figure}

The star formation and stellar mass of NGC~895 were obtained by a multi-wavelength SED fitting analysis exploiting the spectral energy distribution (SED) fitting code \texttt{CIGALE} (v. 2022.1. \citealt{2019A&A...622A.103B}, as applied to multi-wavelength photometry by \citealt{2018A&A...620A..50M}). In order to perform the fit, it is important to have access to the best multi-wavelength compilation of aperture-matched photometry. For this reason, we used the multi-wavelength matched photometry available from DustPedia \citep{2018A&A...620A.112B} combined with our MeerKAT measurements and VLA radio continuum flux measurements. The best fit model is shown in Fig.~\ref{fig:SED}. This was obtained by assuming the following models and parameters. We used a Chabrier initial mass function (IMF) with delayed star formation history and nebular emission (we also tried fitting the SED with an additional star formation burst, but our results did not change, so an additional burst is not needed to explain the results); we used the standard attenuation law presented in \cite{2000ApJ...539..718C} with the interstellar medium (ISM) and birth cloud slope of the attenuation equal to -0.7. We used the modified dust template by \cite{2007ApJ...657..810D}. CIGALE allows performing the fit by exploiting different SED templates and emission models, e.g. AGN torus emission. We performed a run with the AGN module, but the estimated AGN fraction, defined as a ratio of the AGN luminosity to the sum of the AGN and dust luminosities, was lower than 0.3$\%$ (0.24$\pm$0.42). The probability distribution function characterized by a single peak around the AGN fraction equal to 0. It means that even with an extensive range of possibilities for the AGN module, CIGALE did not find even one solution with a significant AGN contribution larger than 0.5$\%$. In the broadband spectral energy distribution fitting, the AGN fraction lower than 15$\%$ is usually too uncertain to be considered a real AGN contribution. Moreover, the goodness of the fit, calculated as a pseudo reduced $\chi^2$, is slightly larger when the AGN module is forced to be used, even with such a small percentage (the reduced $\chi^2$ increases from 1.7 without the AGN module, to 1.8 with AGN module, and AGN fraction lower than 0.2$\%$). Additionally, the stellar masses calculated with and without AGN module agree up to the second decimal (M$_{\star} = 2.28 \pm$ 0.16 $\times 10^{10}$ M$_{\odot}$, and M$_{\star}  = 2.27 \pm$ 0.16 $\times 10^{10}$ M$_{\odot}$ 
for a run without and with the AGN module, respectively). We found no changes in the estimation of the star formation rate of 1.75 $\pm$ 0.09 M$_{\odot}$/yr.
Our finding suggests that this galaxy does not host an AGN, and all the emission is due to star formation activity \citep{1998AJ....115.1693C}.

\begin{figure*}
    \centering
    \includegraphics[width=11.8cm]{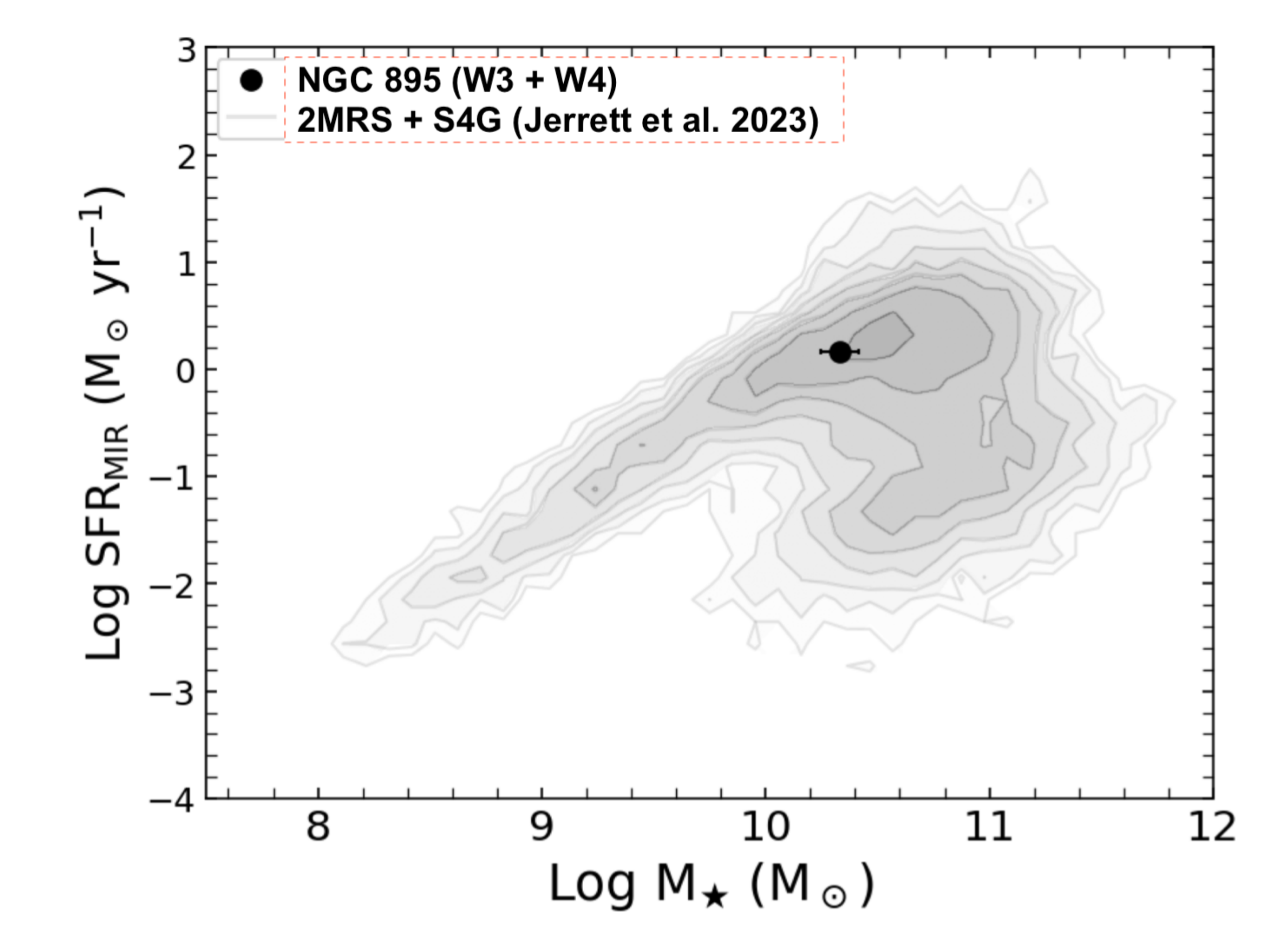}
    \caption{Comparison to main sequence relations. The grey line shows the galaxy sample from  2MRS and S4G and the black circle represents the location of NGC~895 galaxy on the main sequence \citep{2023arXiv230105952J}.}
    \label{sfloc}
\end{figure*}

A similar value for the SFR, of $1.50 \pm 0.14$ M$_{\odot}$/yr 
\citep{2023arXiv230105952J}, was determined using the new scaling relations between WISE mid-infrared galaxy photometry and well determined stellar masses from SED modeling optical-infrared photometry provided by the Galaxy And Mass Assembly (GAMA) Survey\footnote{http://www.gama-survey.org/} data release 4 (DR4 Catalogue of the GAMA-KiDS-VIKING survey of the southern G23 field \citep{2022MNRAS.513..439D}). Fig.~\ref{sfloc} shows the relationship between the star formation rate and the stellar mass for galaxies obtained by \citet{2023arXiv230105952J}. The grey line shows the galaxy sample from the 2MASS redshift survey (2MRS, \citealt{2012ApJS..199...26H}) and the Spitzer Survey of Stellar Structure in Galaxies (S4G, \citealt{2014ApJ...788..144M}). The black circle indicates the location of NGC~895 with respect to the main sequence. NGC~895 lies in the middle of the main sequence with no excess star fomation.

\section{Analysis of the \HI\ -detected satellites}\label{sec:optical}

\subsection{\HI\ properties}

The \HI\ morphology of the two galaxies is irregular. Using the faintest \HI\ contour (see Fig.~\ref{fig:mom0}), we estimate \HI\ diameters of 12.3 $\pm$ 0.9 kpc for J022138 and 8.9 $\pm$ 0.9 kpc for J022042 (the error being defined by the beam size). The profile widths at 50$\%$ levels $\Delta_{50}$ are 52.9 $\pm$ 22.0 \kms and  48 $\pm$ 22.0 \kms for J022138 and J022042, respectively. We derive \HI\ masses of 5.4 $\pm$ 0.3 $\times$ 10$^{7}$ M$_{\odot}$ and 8.2 $\pm$ 0.1 $\times$ 10$^{7}$ M$_{\odot}$ for J022138 and J022042. The \HI\ properties of the companion galaxies are provided in Table \ref{tab:Properties}.

Given the coarse velocity resolution of our data, the size of the companion galaxies, with only two \HI\ beams across, does not allow us to derive \HI\ rotation velocities using kinematic analysis, hence we estimate the maximum rotation velocities using $V_{\text{rot}} = W_{50}/2\text{sin}(i)$, where $W_{50}$ is the width of the \HI\ spectrum at the level of 50$\%$ of the peak value and $i$ is the inclination. The inclination angles were defined as cos$^2(i) = (d^2 - \theta_{d}^2$)/($D^2 - \theta_{D}^2$), where d and D are the minor and major axis measured to the outer contour of the \HI\ moment 0 map and $\theta_{d}$ and $\theta_{D}$ are the sizes of the synthesized beam in the direction of the major and minor axis of the \HI\ disk.  $\theta_{d}$ and $\theta_{D}$ are used to take into account the beam-smearing effect due to our low spatial resolution data. Using the \HI\ parameters in Table \ref{tab:Properties}, we derive the rotation velocities of 38 $\pm$ 22 \kms and 43 $\pm$ 22 \kms for J022138 and J022042, respectively. The dynamical masses are estimated using
\begin{equation} \label{eq:2}
M_{\text{dyn}} =  \frac{R_{\text{\HI}} V_{\text{rot}}^2} {\text{kpc}(\text{km s}^{-1})^2} \times 2.33 \times 10^{5} M_{\odot},
\end{equation}
where V$_{\text{rot}}$ is the rotation velocity. This gives us dynamical masses of 2.0 $\pm$ 0.2 $\times$ 10$^{9}$ M$_{\odot}$ for J022138, and 1.6 $\pm$ 0.1 $\times$ 10$^{9}$ M$_{\odot}$ for J022042.

\subsection{Optical properties}
We analyse the deep optical images of the \HI-detected satellites discussed previously: J022138 and J022042. Photometry in all the available optical bands, and structural properties when possible, are provided in Table \ref{tab:Properties}.

\begin{figure*}
	\includegraphics[width=1.0\textwidth]{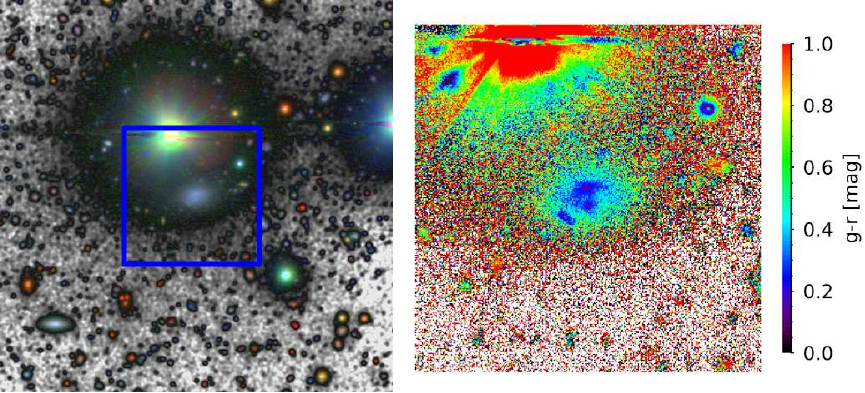}
\caption{Colour composite image of J022042 with \textit{g, r} and \textit{i} bands from HSC-SSP is shown in the left panel. The blue rectangle indicates the zoom-in region shown in the right panel, in which the \textit{g-r} photometric colour is shown.}
    \label{fig:Satellite_small}
\end{figure*}

\begin{table}
\begin{center}
\caption{Summary of properties for J022138 and J022042. Photometric magnitudes are corrected for extinction \citep{2011ApJ...737..103S}.  
Stellar masses and star formation rates are calculated via SED fitting with \texttt{CIGALE}. The $i$ and $y$ magnitudes values reported with * are from the HELP DR1 Merged Catalogue while the $W4$ measurement with no error is to be considered as an upper limit. Magnitudes with $\dag$ are from the HSC-SSP. Magnitudes with $\ddag$ are from DECALS (see text).
\HI\ properties: $^{a}$ \citep{2021A&A...646A..35M}, $^{b}$ (this work)}
\normalsize
\begin{tabular}{c@{\hspace{0.008cm}}c@{\hspace{0.04cm}}c@{\hspace{0.04cm}}}
\label{tab:Properties}
Parameter & J022138 & J022042\\
\hline
RA & 02$^{h}$ 21$^{m}$ 38.07$^{s}$ & 02$^{h}$ 20$^{m}$ 42.1$^{s}$\\
Dec & $-$05\textdegree {} 26' 31.5'' & $-$05\textdegree {} 21' 15.00''\\
\textit{g}-band & 17.13 $\pm$ 0.01 mag\dag& 19.30 $\pm$ 0.18 mag\ddag\\
\textit{r}-band & 16.79 $\pm$ 0.01 mag\dag& 18.93 $\pm$ 0.15 mag\ddag\\
\textit{i}-band & 16.66 $\pm$ 0.01 mag\dag&18.56 $\pm$ 0.01 mag* \\
\textit{z}-band & 16.59 $\pm$ 0.01 mag\dag& 18.68 $\pm$ 0.13 mag\ddag\\
\textit{y}-band & 16.57 $\pm$ 0.02 mag\dag& 21.07 $\pm$ 0.11 mag* \\
W1-band & 17.34 $\pm$ 0.03 mag & - \\
W2-band & 17.89 $\pm$ 0.13 mag & - \\
W3-band & 17.46 $\pm$ 0.29 mag & - \\
W4-band & $>$ 15.31 mag & - \\
Stellar mass & 1.38 $\pm$ 0.12 $\times$ 10$^{8}$ M$_{\odot}$ & 4.21 $\pm$ 0.97 $\times$ 10$^{7}$ M$_{\odot}$ \\
SFR & 0.00403 $\pm$ 0.001 M$_{\odot}$/yr & -\\
$\mu _g$(0) & 22.9 $\pm$ 0.1 mag/arcsec$^2$& 23.4 $\pm$ 0.2 mag/arcsec$^2$\\
S\'ersic index & 0.73 $\pm$ 0.05& 0.83 $\pm$ 0.10\\
Axis ratio & 0.49 $\pm$ 0.01 & 0.71 $\pm$ 0.05\\
{Position angle} & 49 $\pm$ 1.1 degrees& 110 $\pm$ 1.0 degrees \\
Effective radius  & 11.7 $\pm$ 0.4 arcsec&  4.0 $\pm$ 0.5 arcsec \\
\hline
\multicolumn{3}{c}{\HI\ properties} \\
Redshift$^{a}$&0.006912 $\pm$ 0.000006& 0.007359 $\pm$ 0.00007 \\
Total \HI\ mass $^{a}$&5.4 $\pm$ 0.3 $\times$ 10$^{7}$ M$_{\odot}$  & 8.2 $\pm$ 0.1 $\times$ 10$^{7}$ M$_{\odot}$ \\
Dynamical mass$^{b}$& 2.0 $\pm$ 0.2 $\times$ 10$^{9}$ M$_{\odot}$ & 1.6 $\pm$ 0.1 $\times$ 10$^{9}$ M$_{\odot}$ \\
V$_{\text{heliocentric}}$ $^{b}$ & 2057.3 $\pm$ 22.0 \kms & 2188.2 $\pm$ 22.0 \kms \\
W$_{50}$$^{b}$ & 52.9 $\pm$ 22.0 \kms& 48 $\pm$ 22.0 \kms \\
\HI\ radius$^{b}$ & 6.2 $\pm$ 1.2 kpc & 4.5 $\pm$ 0.5 kpc \\
Inclination$^{b}$ & 43.9 $\pm$ 4.3 degrees & 33.1 $\pm$ 3.3 degrees \\

    \hline
\end{tabular}
\end{center}
\end{table}
First, we analyse J022042. In Fig. \ref{fig:Satellite_small} we show an image with the optical data available in the HSC-SSP dataset. The chance presence of an adjacent star is problematic for the analysis of the lower surface brightness regions in J022042 and problematic in general for a photometric analysis. To obtain cleaner photometry, we use DECALS data in which stars are modelled and subtracted from the images, allowing the effect of the adjacent star to be significantly reduced. This allows us to perform aperture photometry, which is presented in Table \ref{tab:Properties}.

Although an analysis of low surface brightness regions, for example with photometric profiles, is not possible for J022042 due to the contamination of an adjacent star, we use the HSC-SSP images of high depth and resolution  to analyse the substructure. In the right panel of Fig. \ref{fig:Satellite_small}, we present a \textit{g-r} colour map which shows the presence of clumps of star formation in the central region of J022042, specifically a central clump of larger extension and a smaller one in the NW direction. While the integrated colour of J022042 is \textit{g-r} = 0.37 mag (see Table \ref{tab:Properties}), the clumps have a bluer colour of approximately \textit{g-r}~$\sim$~0.2~-~0.3~mag. Regarding the morphology of J022042, just attending to the visual morphology in the color map, we did not find significant asymmetries beyond the presence of these clumps of star formation in the central regions. The main body of this satellite appears redder in colour and quite symmetrical.

\begin{figure*}
	\includegraphics[width=1.0\textwidth]{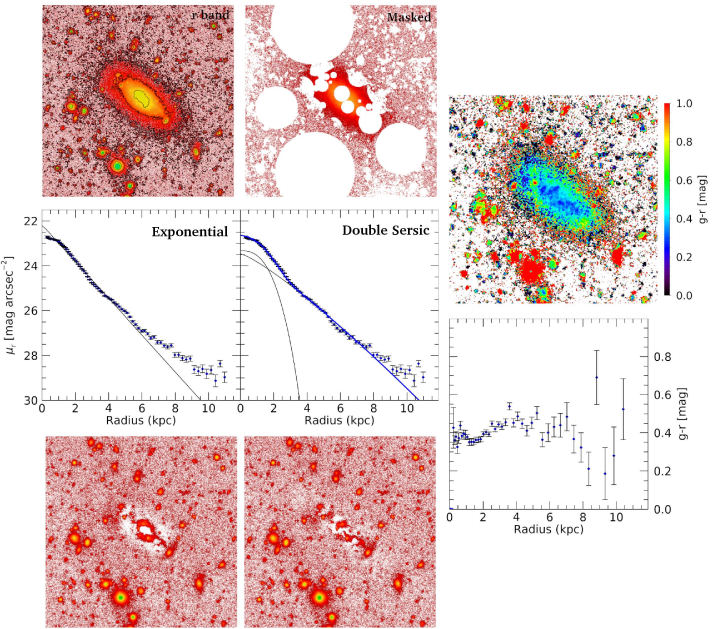}
\caption{Optical analysis of J022138. The upper left panels show the r-band image with isophotes at 23, 25 and 27 mag arcsec$^{-2}$ and the masked image used to derive the profiles. In the middle left panels the profiles and the fit with an exponential and double Sersic models are shown as solid black lines. The solid blue line indicates the sum of the individual profiles in the double Sersic modelling. The observed profile of J022138 is shown as points with error bars. In the lower left panels, the residuals after the subtraction of the models are shown. The upper right panel is a g-r colour image. The lower right panel shows the g-r colour profile for J022138 with its error bars.}
    \label{fig:Satellite}
\end{figure*}

We now focus on J022138. The larger extension of J022138 and the absence of bright foreground sources allows us to make a much more detailed study of this satellite. We show a summary of its properties in Table \ref{tab:Properties}. In Fig. \ref{fig:Satellite} we show the results of an analysis of its photometric profiles. First, we find that a single Sersic model is not a good representation of the morphology of the galaxy. Both fitting a single Sersic model with free Sersic index n and fixing to an exponential profile n=1 produce an excess of light in the outer parts (shown in Fig. \ref{fig:Satellite}). This is also visible in the residual image. The fit with two Sersic models shows a much better fit. We can see in Fig. \ref{fig:Satellite} that the profile is well represented by this double Sersic model, showing a much cleaner residual image. Even in the best fit, an excess of light in the outer parts, starting at about 9\,kpc still remains in the profile.
To further investigate the double Sersic model, we show in Fig. \ref{fig:Satellite} a panel with the photometric colour g-r of the satellite together with a photometric profile of this colour. The innermost region, approximately the inner 2\,kpc, has a bluer colour than the rest of the satellite where we can also observe HII regions of star formation. We conclude that the central region is dominated by star formation, producing an artificial second component in the photometric profile.

We also show the isophotal contours of the satellite at 23, 25 and 27 mag arcsec$^{-2}$ in Fig. \ref{fig:Satellite}. While the inner regions adjust to a more elliptical morphology, the contour of 27 mag arcsec$^{-2}$ tends towards a boxy shape. The excess of light in the external regions observed in the profile together with their boxy morphology, could be an indication of a minor tidal interaction. We disregard problems derived from the sky subtraction due to the absence of over-subtraction at large radius.

\begin{figure}
    \centering
    \includegraphics[width=8.5cm]{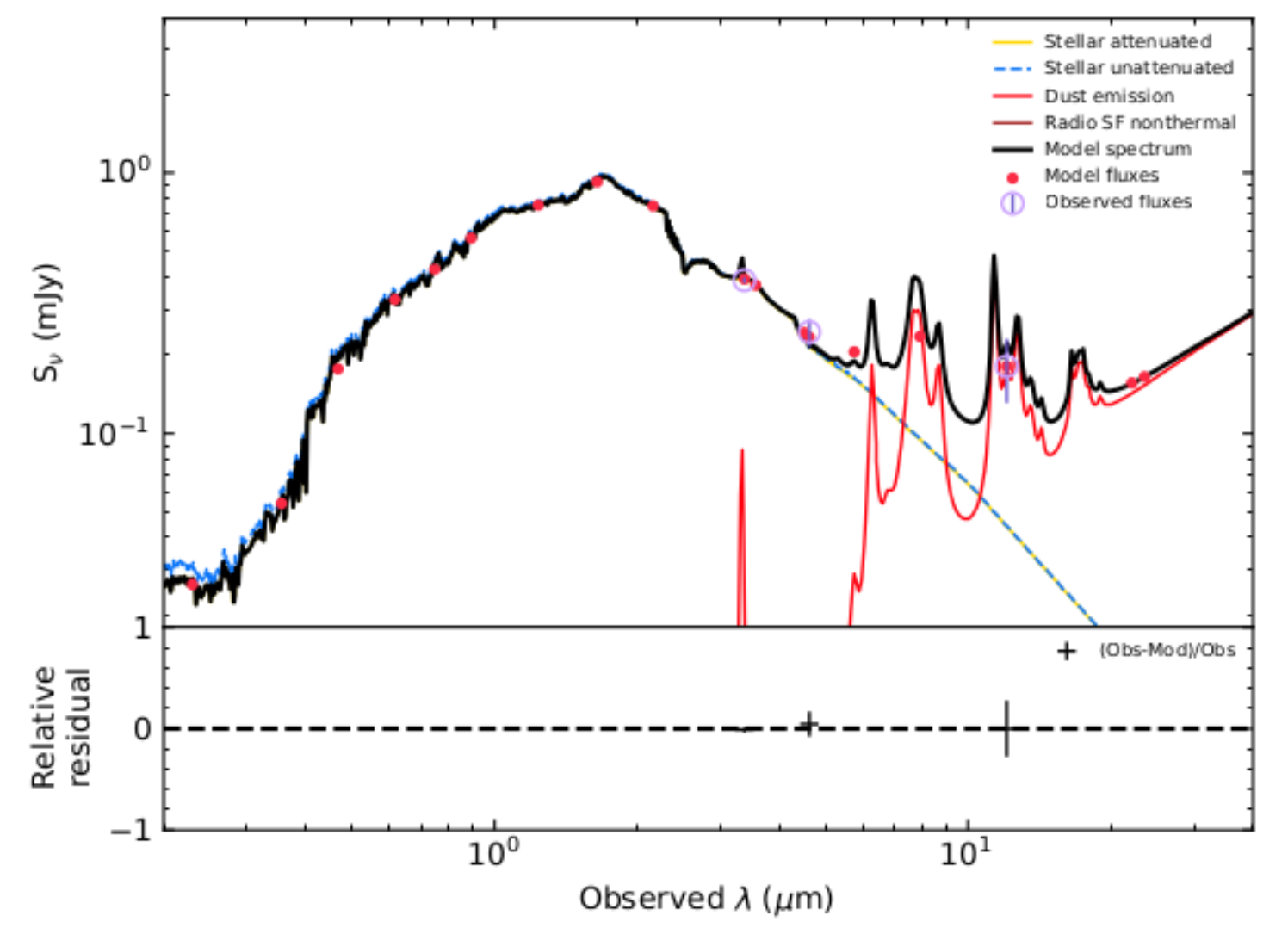}
    \caption{\texttt{CIGALE} SED Fitting results of J022138. In the case of J022138, we did not plot the nebular emission for the clarity of the figure. The reduced  $\chi^2$ of the fit is equal to 1.7.}
    \label{fig:SED1}
\end{figure}

We used \texttt{CIGALE} to perform the SED fitting of both satellite galaxies and thus obtain an estimate of their stellar masses and star formation rates. In order to run the fit we exploited the optical and WISE photometry reported in Table \ref{tab:Properties}. The WISE photometric measurements were carried out using the techniques developed by \cite{2019ApJS..245...25J} but are only available for J022138 as J022042 could not be reliably measured in WISE images. We complemented our optical photometric measurements against the photometry provided by the Herschel Extragalactic Legacy Project \citep[HELP]{2016ASSP...42...71V,2019MNRAS.490..634S,2021MNRAS.507..129S} DR1 Merged Catalogue\footnote{\url{https://hedam.lam.fr/HELP/}} and found that the two photometric datasets were in good agreement for the bands we could measure ourselves, with HELP providing a few extra data points. The best fit model for J022138 is shown in Fig.~\ref{fig:SED1}. We measure the star formation rate of 0.004 $\pm$ 0.001 M$_{\odot}$/yr for J022138. This value is comparable to those of isolated dwarf irregular galaxies with typical star formation rates, such as WLM (0.006 M$_{\odot}$/yr), IC 1574 (0.005 M$_{\odot}$/yr), and AndIV (0.003 M$_{\odot}$/yr) \citep{2009ApJ...706..599L}. Therefore, we do not observe any star formation enhancement in  J022138. However, even including the HELP measurements, we could not strongly constrain the long-wavelength SED of J022042 due to the lack of a WISE detection, so for this object we could only estimate the stellar mass but not the star formation rate. The derived values are reported in Table \ref{tab:Properties}.

\section{Discussion}\label{sec:discuss}
 One of the main questions emerging from the data presented is the possible origin of the \HI\ features observed in NGC~895. Asymmetries are common in disc galaxies. They are often seen in the form of morphological, kinematical or integrated \HI\ spectral asymmetries \citep{2008A&ARv..15..189S, 2009PhR...471...75J,2011A&A...532A.117E, 2014A&A...569A..68D}. Among the different mechanisms used to explain such asymmetries in galaxies are tidal interactions. Here, we discuss the implications of the above results by considering several possible interaction scenarios that may have caused the observed \HI\ and optical features detected in NGC~895.
 
\subsection{Environment and morphology of NGC~895}
We search for signatures of interaction by examining the environment around NGC~895 and its companion galaxies as well as their \HI\ morphologies (see Fig.~\ref{fig:mom0}). Our deep \HI\ observations reveal several features in NGC~895 that are clear signatures of interactions. The presence of symmetric tidal arms, as well as the lopsided and warped structure and kinematics of NGC~895 are suggestive of the strong effects of interaction with its environment. The distribution of the \HI\ emission is asymmetric with more \HI\ emission towards the receding side as compared to the approaching side (see Fig.~\ref{fig:prof}). 
Several studies have demonstrated that warping and lopsidedness of mass distribution can be observed in a significant fraction of galaxies \citep{2008A&ARv..15..189S,1998AJ....115...62H,2009A&A...494..489J}. The presence of these features is generally interpreted as a sign of interaction. Given that these features are visible in NGC 895, it is important to know how they originated.

Although previously identified as an isolated galaxy in \HI\ by \citet{1988ngc..book.....T}, it is immediately obvious from Fig.~\ref{fig:mom0} that NGC~895 has 2 gas-rich dwarf companions located north and north-west of the main galaxy. These newly detected \HI\ companions are identified as J022138 and J022042. Signatures of interactions between interacting galaxies can often be seen in the form of bridges or tidal tails between galaxies \citep{2013ApJ...779L..15N}. While bridges are associated with early stages of ongoing interactions, tidal tails between galaxies are often linked to past close encounter events with other galaxies. We searched for tidal tails and bridges between NGC~895 and its companions, at column densities of $\sim$ 10$^{20}$ cm$^{-2}$, and find no evidence for tidal tails and bridges. Although \HI\ tidal tails and bridges have been detected at this column density before \citep{2018ApJ...865...26D, 2019MNRAS.487.5248L, 2021MNRAS.505.3795N}, it is likely that the diffuse \HI\ debris has a lower \HI\ column density ($\leq$ 10$^{20}$ cm$^{-2}$) that cannot be detected by our current observations. 

Even though the environment around NGC~895 does not show significant signs of interactions such as tidal tails and bridges, we estimate the upper limit orbital period of each companion galaxy, which is the amount of time it will take for a companion to orbit around the centre of the main galaxy and compare these values to the age of the tails associated with NGC~895. The orbital period is 2$\pi$R/V, where R is the separation between the centres of the main galaxy and companion in kpc, and V is the velocity difference in \kms. We estimate upper limit orbital periods of $\sim$ 1\,Gyr and 8\,Gyr for J022138 and J022042, respectively. Given the length of the western tail ($\sim$ 37\,kpc) and the mean velocity dispersion of NGC~895 (16.8 \kms), the age  (T = length/dispersion) of the tail must be at least of order 2.2\,Gyr. This suggests that it is unlikely that J022042 has caused the asymmetries associated with NGC~895. However, looking at J022138 with an orbital period of 1\,Gyr, a possible scenario is that J022138 did a close fly-by of NGC~895, warping the \HI\ disc of the central galaxy, while disrupting its own stellar and \HI\ distribution. Such a scenario is supported by the excess of light in the external regions observed in the profile which together with their boxy morphology in Fig. \ref{fig:Satellite}, could be an indication of tidal interaction \citep{1999ApJ...523L.133N}.

\subsection{Tidal stripping}
\HI\ tails and streams, as well as stellar streams, between galaxies are often associated with tidal stripping of the smaller companion by the larger galaxy (for example \citealt{2021ApJ...913...53P,2021ApJ...921L..36L,2014MNRAS.438.1435C,1998Natur.394..752P}). The deviation from a Sersic model through an excess of light in the outer parts could be considered an indication of the external perturbation. In the deep optical images, we do not find any evidence for a tidal tail or stream. To further investigate the potential interactions that may take place, we estimate the tidal radius, (r$_{\text{tidal}}$), the radius beyond which matter is tidally stripped due to an interaction between two galaxies, given by \cite{1984AJ.....89..966D} as:
\begin{equation}
r_{\text{tidal}} \geq D \times \big[\frac{M_{\text{companion, dyn}}}{M_{895}, \text{dyn}}\big]^{1/3},
\end{equation}
where $D$ is the radially projected distance between the NGC 895 and the companion, while $M_{\rm \text{companion},dyn}$ and $M_{\rm \text{895},dyn}$ are the dynamical masses for the \HI\ companions and NGC~895 respectively.We derive the r$_{\text{tidal}}$ of 10.3 and 34.3\,kpc for J022138 and J022042 respectively. The calculated tidal radius of each satellite is much larger than their \HI\ radii of 6.2\,kpc and 4.5\,kpc, meaning that it is unlikely that they are prone to disruption.

Whereas for J022042 the tidal radius is much larger than the extension of the galaxy, for J022138 we observe that r$_{\rm \text{tidal}}$ coincides approximately with the region in the profile in which we observe an excess of light. This could indicate that, although we cannot detect the presence of tidal tails/streams, probably due to insufficient depth in the optical data, we detect the tidal effects produced by NGC~895 on the satellite, and therefore uncover evidence for an interaction between the two systems. We note that the estimate of the tidal radius is derived using the current projected distance between the main galaxy and its companion. In the past, the galaxies could have been closer to each other.

\subsection{Link with star formation}
The occurrence of high SFR in galaxies has been associated with galaxy interactions \citep{1994Natur.372..530Y}. Previous studies have found that mergers and interactions can enhance the SF in galaxies which may lead to starburst episodes (e.g. \citealt{2015MNRAS.454.1742K, 2018ApJS..234...35Z,2009ApJ...698.1437K, 2000ApJ...530..660B,2020A&A...635A.197D, 2021ApJ...909..120S}).
The fact that NGC~895 lies on the main sequence is a relatively strong indication that no noticeable SFR enhancement has occurred. Our analysis shows that if there have been interactions of NGC ~895 with its environment, these have not enhanced its SFR. This is in agreement with \citet{1984PASP...96..273B} who showed that even though galaxies in their sample were involved in strong interactions, it was not immediately obvious that the central regions had been greatly affected as their calculated SFR for most interacting galaxies fell in the near constant SFR regime. The implication of this for our analysis is that the typical value of SFR calculated for NGC ~895 might not be sufficient to rule out possible interactions.

\section{Conclusions}\label{sec:conclude}
We present high-sensitivity MIGHTEE-\HI\ observations of NGC~895 as well as deep archival HSC optical images. We highlight the results of this study:

\begin{enumerate}

\item The \HI\ morphology and kinematics of NGC~895 are asymmetric. The \HI\ profile has more \HI\ emission from the receding side than the approaching side. The \HI\ kinematics shows that the disc of NGC~895 has a warp, which might indicate a disturbance. 

\item Assuming a distance of 34.3 $\pm$ 1.1 Mpc, we derive an \HI\ mass of 1.4 $\pm$ 0.01 $\times$ 10$^{10}$ M$_{\odot}$ for NGC~895, a value consistent with that derived from single-dish observations.

\item At column densities of 1.2 $\times$ 10$^{20}$ cm$^{-2}$, we detect two new \HI\ companions of NGC~895, thus changing the narrative that it is an isolated galaxy. These are classified as MGTH$\_$J022138.1-052631 and MGTH$\_$J022042.1-052115.

\item The \HI\ masses for the companion galaxies are 8.2 $\pm$ 0.1 $\times$ 10$^{7}$ M$_{\odot}$ and 5.4 $\pm$ 0.3 $\times$ 10$^{7}$ M$_{\odot}$ for J022138 and J022042 respectively. 

\item Although the \HI\ properties of NGC~895 indicate that it might have undergone an interaction, we do not find tidal debris between NGC~895 and its companions. This might indicate either that other mechanisms are responsible for the asymmetries seen in NGC~895 or that our observation sensitivity does not allow us to detect these interaction features which could be present at low column density levels. 

\item We calculate the orbital periods of the \HI\ companion galaxies and compare these to the age of the \HI\ tail of NGC~895. We find that it is unlikely that J022042 is responsible for the observed interaction features in NGC~895. It is, however, possible that J022138 did a close fly-by of NGC~895, warping the \HI\ disc of the central galaxy, while disrupting its own stellar and \HI\ distribution. This is also supported by the fact that the calculated tidal radius of each satellite is much larger than their \HI\ radius, meaning that it is unlikely that they are prone to disruption. For J022138, the tidal radius is of the order of the radius at which we observe a slight excess of light in the optical profile and it might well be that during an earlier phase of encounter, when the companion may have been closer to NGC~895, tidal effects did play some role.  

\item We find that the SFR of NGC~895 has not been enhanced. The galaxy has an active star-forming disk and lies on the main sequence with a SFR of $\mathrm{SFR =1.75\pm0.09}$ $[M_{\odot}/yr]\ $. Based on previous studies, this does not rule out any possible interaction with its companion galaxies.

Our work shows the high potential and synergy of using state-of-the-art data in both \HI\ and optical to reveal a more complete picture of galaxy environments.

\end{enumerate}

\section*{Acknowledgements}
The MeerKAT telescope is operated by the South African Radio Astronomy Observatory, which is a facility of the National Research Foundation, an agency of the Department of Science and Innovation. BN's research is supported by the South African Radio Astronomy Observatory (SARAO). We acknowledge the use of the ilifu cloud computing facility - \url{www.ilifu.ac.za}, a partnership between the University of Cape Town, the University of the Western Cape, the University of Stellenbosch, Sol Plaatje University, the Cape Peninsula University of Technology and the South African Radio Astronomy Observatory. The Ilifu facility is supported by contributions from the Inter-University Institute for Data Intensive Astronomy (IDIA – a partnership between the University of Cape Town, the University of Pretoria, the University of the Western Cape and the South African Radio astronomy Observatory), the Computational Biology division at UCT and the Data Intensive Research Initiative of South Africa (DIRISA). BN acknowledges financial support from the Women by Science programme of the Fundaci\'on Mujeres por \'Africa, from the Gobierno de Canarias, and from the Instituto de Astrof\'\i sica de Canarias (IAC). JR and JHK acknowledge support from the Spanish Ministry of Science and Innovation under the grant "The structure and evolution of galaxies and their central regions" with reference PID2019-105602GB-I00/10.13039/501100011033, from the ACIISI, Consejer\'{i}a de Econom\'{i}a, Conocimiento y Empleo del Gobierno de Canarias and the European Regional Development Fund (ERDF) under grant with reference PROID2021010044, and from IAC project P/300724, financed by the Ministry of Science and Innovation, through the State Budget and by the Canary Islands Department of Economy, Knowledge and Employment, through the Regional Budget of the Autonomous Community. JR also acknowledges funding from University of La Laguna through the Margarita Salas Program from the Spanish Ministry of Universities ref. UNI/551/2021-May 26, and under the EU Next Generation. JF-B  acknowledges support through the RAVET project by the grant PID2019-107427GB-C32 from the Spanish Ministry of Science, Innovation and Universities (MCIU), and through the IAC project TRACES which is partially supported through the state budget and the regional budget of the Consejer\'ia de Econom\'ia, Industria, Comercio y Conocimiento of the Canary Islands Autonomous Community. RI acknowledges financial support from the grant CEX2021-001131-S funded by MCIN/AEI/10.13039/501100011033, from the grant IAA4SKA (Ref. R18-RT-3082) from the Economic Transformation, Industry, Knowledge and Universities Council of the Regional Government of Andalusia and the European Regional Development Fund from the European Union and financial support from the grant PID2021-123930OB-C21 funded by MCIN/AEI/10.13039/501100011033, by "ERDF A way of making Europe" and by the "European Union" and the Spanish Prototype of an SRC (SPSRC) service and support funded by the Spanish Ministry of Science and Innovation (MCIN), by the Regional Government of Andalusia and by the European Regional Development Fund (ERDF). EN and THJ acknowledge support from the National Research Foundation (South Africa). OS's research is supported by the South African Research Chairs Initiative of the Department of Science and Technology and National Research Foundation (grant number 81737). LM and MV acknowledge financial support from the Inter-University Institute for Data Intensive Astronomy (IDIA), a partnership of the University of Cape Town, the University of Pretoria, the University of the Western Cape and the South African Radio Astronomy Observatory, and from the South African Department of Science and Innovation's National Research Foundation under the ISARP RADIOSKY2020 Joint Research Scheme (DSI-NRF Grant Number 113121) and the CSUR HIPPO Project (DSI-NRF Grant Number 121291). SHAR is supported by the South African Research Chairs Initiative of the Department of Science and Technology and National Research Foundation. KM is grateful for support from the Polish National Science Centre via grant UMO-2018/30/E/ST9/00082. NM acknowledges support of the LMU Faculty of Physics.MJJ, IH and AAP  acknowledge support of the STFC consolidated grant [ST/S000488/1] and  [ST/W000903/1] and MJJ  and IH support from a UKRI Frontiers Research Grant [EP/X026639/1]. MJJ also  acknowledges support from the Oxford Hintze Centre for Astrophysical Surveys which is funded through generous support from the Hintze Family Charitable Foundation. LVM acknowledges financial support from the grant SEV-2017-0709 funded by MCIN/AEI/ 10.13039/501100011033, from the grants RTI2018-096228-B-C31 and PID2021-123930OB-C21 funded by MCIN/AEI/ 10.13039/501100011033, by “ERDF A way of making Europe” and by the "European Union" and from the grant IAA4SKA (R18-RT-3082) funded by the Economic Transformation, Industry, Knowledge and Universities Council of the Regional Government of Andalusia and the European Regional Development Fund from the European Union.

\section*{Data availability}
The \HI\ data from this study are available as part of the first MIGHTEE survey data release and the optical data were publicly available.


\bsp
\label{lastpage}
\end{document}